\documentclass[11pt]{elsarticle}

\usepackage{amsmath,amssymb,amsfonts}
\usepackage{graphicx}
\usepackage{hyperref}
\usepackage{booktabs}
\usepackage{braket}
\usepackage{amsthm}

\journal{Applied Mathematics and Computation}

\theoremstyle{plain}
\newtheorem{theorem}{Theorem}
\newtheorem{lemma}{Lemma}
\newtheorem{proposition}{Proposition}
\newtheorem{corollary}{Corollary}

\newtheorem{remark}{Remark}

\usepackage{algorithm}
\usepackage{algpseudocode}

\newcommand{\erf}[1]{\ensuremath{ \operatorname{erf} \left(  #1 \right)   } }
\newcommand{\erfc}[1]{\ensuremath{ \operatorname{erfc} \left(  #1 \right)   } }

\newcommand{\hsym}[2]{  \ensuremath{ \,{}_{#1}F_{#2}\! }  } 
\newcommand{\fclass}[2]{\ensuremath{  \mathbb{#1}^{\, #2} }}

\begin{document}
	
	\begin{frontmatter}
		
		\title{Hybrid Fredholm-Variational Method for Eigenvalue Problems with Error-Function Screened Coulomb Potentials}
		
		\author[inst1]{Dimiter Prodanov\corref{cor1}}
		\ead{dimiter.prodanovl@iict.bas.bg}
		
		\cortext[cor1]{Corresponding author}
		\address[inst1]{Laboratory of Neurotechnology, PAML-LN, IICT, BAS, Sofia, Bulgaria}
		
	\begin{abstract}
		This work reformulates the ground-state problem for the error-function screened Coulomb potential, the
		long-range part of the Coulomb operator in range-separated hybrid functionals, as a Fredholm
		integral equation built on the free Green's function, and combines it with variational optimization
		of a single decay parameter. This gives an analytic route to the ground state: the first
		approximation to the wavefunction is expressed in closed form through standard special functions,
		and every optimized energy is a rigorous upper bound. We prove that the iteration converges
		geometrically to the ground state at the exact decay parameter, derive a priori bounds on the
		energy and on the optimal parameter, and control the slope of the wavefunction at the origin. A
		harmonic-oscillator homotopy supplies a further upper bound that is informative only at large
		screening length. Over a wide range of screening lengths the first two approximations reproduce
		the exact energy to within a fraction of a percent. . 
		Numerical results are compared with shooting, Numerov matching and Gaussian basis sets. The
		closed-form iterates are most economical in the Coulomb-like regime of small screening length,
		where a one-parameter trial function matches the accuracy of several Gaussians, whereas for large
		screening length the nearly harmonic potential favours Gaussian bases. The analytic structure also
		supplies initial data and rigorous brackets for standard solvers.

	\end{abstract}
		
	\end{frontmatter}
	
	\section{Introduction}\label{sec:intro}
	
	Screened Coulomb potentials arise in plasma physics, solid-state theory and quantum
	chemistry~\cite{ichimaru1982,dehesa1987}. The error-function screened potential
	$V_\sigma(r)=-\operatorname{erf}(r/2\sigma)/r$ is the long-range part of the Coulomb operator
	in the range-separated hybrid functionals of the HSE family of density functional
	theory~\cite{Heyd2003,Heyd2006}. It is regular at the origin, $V_\sigma(0)=-1/(\sigma\sqrt\pi)$,
	and tends to the Coulomb potential as $\sigma\to0$ and to zero as $\sigma\to\infty$. Unlike
	exponentially screened (Yukawa, Debye--H\"uckel) potentials, it has, to our knowledge, received
	little analytical treatment for the bound-state problem.
		
	 For the $s$-wave ground state, which is the subject of this paper, the problem is one-dimensional
	 and several established methods apply. Shooting and Numerov integration of the radial equation,
	 with the energy adjusted until the solution has the correct nodal and asymptotic
	 behavior~\cite{press2007,johnson1978}, determine the energy to high accuracy. Variational
	 expansions in a Gaussian basis, whose matrix elements are analytic for this
	 potential~\cite{landau1977,griffiths2018}, are the standard alternative in electronic-structure
	 practice. Perturbation theory around the harmonic oscillator or the hydrogen
	 atom~\cite{sakurai2017,merzbacher1998} is available in principle, but the screened potential has no
	 small parameter that is uniform in the screening length.
	 
	 What the above non-perturbative methods do not provide is an explicit analytic structure. A principled approach that offers such structure is to write the radial equation as a
	 Fredholm integral equation of the second kind with the free Green's
	 function~\cite{joachain1983,taylor1972}, whose kernel is Hilbert--Schmidt and whose Green's
	 function builds in the exponential decay of a bound state. Iterating the integral operator on a
	 simple exponential starting function, combined with variational optimization of its decay
	 parameter, yields a hierarchy of trial functions with upper-bound energies. 
	 The present work develops
	 that approach for the error-function screened Coulomb potential. The paper is organized as follows. 
	Section~\ref{sec:mathematical-formulation} formulates the integral equation and proves the operator
	bounds. Section~\ref{sec:harmonic-oscillator-homotopy} treats the harmonic-oscillator homotopy and
	Sec.~\ref{sec:convergence} the convergence of the iteration. Sections~\ref{sec:results} and
	\ref{sec:numer} give the numerical results, the algorithms and the comparison with shooting,
	Numerov and Gaussian-basis calculations.
	
	\section{Mathematical Formulation}\label{sec:mathematical-formulation}
	The error-function (erf) screened Coulomb potential
	\begin{equation}\label{eq:erfcoul}
		V(r) = -\frac{\erf{r/2\sigma}}{r}
	\end{equation}
	was introduced in the scope of DFT \cite{Heyd2003,Heyd2006} 
	in the form of the Heyd-Scuseria-Ernzerhof (HSE) screened hybrid density functional as a means to computationally split exchange contributions.
	This potential represents a class of potentials where the screening length $\sigma$ modulates the transition from the pure Coulomb limit to unbound state ($\sigma\to\infty$, $\text{erf}\to 0$). In its original intention the screening parameter was treated as purely computational construct. 
	
	On the other hand, an alternative interpretation can be put forward. 
	The erf-screened potential acquires the role of the effective potential resulting from the Coulomb interaction, generated by a Gaussian charge density distribution.  
	This form arises naturally when the bare point charge center undergoes Brownian motion in a harmonic trap, as the smeared charge density seen by a slower degree of freedom (e.g., an electron) becomes Gaussian.
	Conversely, the complementary erfc-screened short-range operator in HSE functionals, represents a point charge screened by a surrounding fluctuating Gaussian neutralizing shell, modeling environmental screening around a fixed center. 
	Both mechanisms share the same Gaussian statistics but describe opposite physical pictures: the center motion produces erf screening (delocalized source), while the shell motion produces erfc screening (delocalized screening). 
	Mathematically, they satisfy the complementary Poisson equations
	\begin{equation}
		\nabla^2[\erf{r/2\sigma}/r] = -\frac{1}{2 \sqrt{\pi}\sigma^3} e^{-r^2/4\sigma^2}
	\end{equation}
	and 
	\begin{equation}
		\nabla^2[\erfc{r/2\sigma}/r] = - 4 \pi \delta{(r)} + \frac{1}{2\sqrt{\pi} \sigma^3} e^{-r^2/4\sigma^2},
	\end{equation}
	where also $\text{erf} + \text{erfc} = 1 $ thus
	recovering the bare Coulomb limit. 
	In this sense, the screening length $\sigma$ encodes the root-mean-square (RMS) drift due to environmental fluctuations, providing a lumped model of decoherence-induced blurring of the interaction source.

	\subsection{Dimensional Consistency} 
	All quantities in this work are expressed in atomic units, where fundamental constants are set to unity: $\hbar = m_e = e = 4\pi\epsilon_0 = 1$. Consequently, physical quantities become dimensionless numerical values corresponding to natural scales: lengths are measured in Bohr radii ($a_0$), energies in Hartrees ($E_h$), and charges in elementary units. Parameters such as the wavevector $k$ and screening length $\sigma$ thus represent dimensionless ratios $k a_0$ and $\sigma/a_0$, ensuring all analytical expressions are dimensionally consistent and numerically evaluable without additional conversion factors. This convention eliminates explicit physical dimensions from the mathematical formalism while preserving correct scaling relationships.

	\subsection{The Fredholm Integral Equation Approach}
	We consider the  $s$-wave states having radial symmetry ($\ell=0$). 
	The bounded states satisfy the Schr\"{o}dinger equation:
	\[
	\left(-\frac{1}{2}\nabla^2 + V(r)\right)\psi = E\psi, \quad
	V(r) = -\frac{\erf{\frac{r}{2\sigma}}}{r}.
	\]
	Using the standard substitution $\psi(r) = y(r)/r$, we obtain the reduced Schr\"odinger radial equation 
	\begin{equation}\label{eq:reduced}
		-\tfrac12 y'' + V(r) y = E y  
	\end{equation}
	and refer to the function \textit{y} as the reduced wave function.
	We consider homogeneous boundary conditions $y(0) = y(\infty) =0$.
	This  time-independent radial Schr\"odinger equation can be reformulated as a Fredholm integral equation of the second kind using Green's function methods \cite{joachain1983,taylor1972}.
	
	The equation defines a reduced Hamiltonian operator
	\begin{equation}
		H: =  -\tfrac12 \frac{d^2}{dr^2} + V(r)
	\end{equation} 
	which will be discussed throughout the paper.
	Observe also that the kinetic energy term 
	\begin{equation}
		T := -\tfrac12 \int_0^\infty y(r) y''(r) dr
	\end{equation}
	can be brought into the form
	\[
	T = \tfrac12 \int_0^\infty [y'(r)]^2 dr
	\]
	since  boundary terms vanish.
	
	
	For the stationary radial Schr\"odinger equation $ Hy = E y$ with potential $V(r)$, the Fredholm integral equation is formulated as:
	\begin{equation}\label{eq:fredholmg}
		y(r) = \int_0^\infty G(r,r';k) V(r') y(r') \, dr', \quad E=-k^2/2
	\end{equation}
	where the Green's function for the free particle labeled by \textit{k} can be obtained as \cite{stakgold2011,Simon2000}.
\begin{equation}\label{eq:greenfree}
	G_k(r,r') = -\frac{2}{k}\,\sinh(kr_<)\,e^{-kr_>}
	= -\frac{1}{k}\left(e^{-k|r-r'|}-e^{-k(r+r')}\right),
\end{equation}
	with $r_< = \min(r,r')$ and $r_> = \max(r,r')$.
	Equivalently,
	$G_k = - 2 \mathcal{L}^{-1} $ with $\mathcal{L} =- d^2/dr^2 + k^2$. 
	The derivation is standard and is presented in \ref{sec:green}.
	Therefore, 
	\[
	y(r) = \frac{2}{k} \int_0^\infty \frac{\erf{ \frac{r'}{2\sigma}}}{r'} e^{-k r_>} \sinh(k r_<) y(r') \, dr'.
	\]
	so that the equation splits in  "retarding" and "advancing" terms as follows 
	\begin{multline}\label{eq:fredholm}
		y(r) = \frac{2}{k} e^{-k r} \int_0^r \frac{\erf{ \frac{r'}{2\sigma} }}{r'} \sinh(k r') y(r') \, dr' \\
		+ \frac{2}{k} \sinh{(k r)} \int_r^\infty \frac{\erf{ \frac{r'}{2\sigma} }}{r'} e^{-k r'} y(r') \, dr' \\ 
		= \frac{2}{k} e^{-k r} J_1(r) + \frac{2}{k} \sinh{(k r)} J_2 (r)
	\end{multline}
	where $J_{1}$ is the retarding and $J_{2}$ is the advancing integral term.
	Note that this function should be further normalized.
	
	
	The point spectrum of $H$ is in one-to-one correspondence with the unit-eigenvalue
	spectrum of the compact Fredholm operator $K(E)$, justifying the
	spectral analysis and iteration on $K$ as an exact reformulation of the original
	Schr\"odinger eigenvalue problem. 
	This technical result is established in \ref{sec:proofL2}. 
	Therefore, it suffices to establish the properties of the operator $K$.

	\subsection{Operator Norm Bound}
	
	\begin{theorem}[Fredholm Operator Norm Bound]\label{th:bond}
		Let $K_k$ be the integral operator on $L^2(0,\infty)$ defined by
		\begin{equation}
			(K_k f)(r) := \int_0^\infty K_k(r,r')\, f(r') \, dr',
		\end{equation}
		where the kernel is
		\begin{equation}\label{eq:KernelK}
			K_k(r,r') := G_k(r,r')\,V(r') = \frac{2}{k}\,\frac{\erf{r'/2\sigma}}{r'}\,\sinh(kr_<)e^{-kr_>}
		\end{equation}
		$V(r) = -\erf{r/2\sigma}/r$, and $G_k$ is the free-particle Green's function \eqref{eq:greenfree} at energy $E = -k^2/2$. 
		Then for all $k>0$ and $\sigma>0$,
		\begin{equation}
			\|K_k\|_{L^2(0,\infty)} \le \frac{2}{\sqrt{\pi}\,\sigma\, k^2}.
		\end{equation}
	\end{theorem}
	
	\begin{proof}
		The proof proceeds in several steps.
		
		\noindent\textbf{Step 1: Operator factorization.}
		
		Define the multiplication operator $M$ and the integral operator $S$ by
		\begin{align}
			(Mf)(r) &:= m(r) f(r), \qquad m(r) := \frac{\erf{r/2\sigma}}{r}, \\
			(Sf)(r) &:= \frac{1}{k}\int_0^\infty \left(e^{-k|r-r'|} - e^{-k(r+r')}\right) f(r') \, dr'.
		\end{align}
		Then the kernel of $K_k$ is $K_k(r,r')=S(r,r')\,m(r')$, so that
		\begin{equation}
			(K_k f)(r) = \int_0^\infty S(r,r')\,m(r')\,f(r')\,dr' = (S M f)(r),
		\end{equation}
		i.e.\ $K_k = S M$ (not $MS$).
		
		By submultiplicativity of the operator norm,
		\begin{equation}
			\|K_k\| \le \|S\|\, \|M\|.
		\end{equation}
		
		\noindent\textbf{Step 2: Norm of the multiplication operator $M$.}
		
		For a multiplication operator on $L^2(0,\infty)$,
		\begin{equation}
			\|M\| = \sup_{r>0} |m(r)| = \sup_{r>0} \frac{\erf{r/2\sigma}}{r}.
		\end{equation}
		
		Let $a := 1/(2\sigma) > 0$. Then $m(r) = \erf{ a r}/r$.
		
		First, note the asymptotic behavior:
		\begin{itemize}
			\item As $r \to 0^+$: $\erf{ar} \sim \frac{2a}{\sqrt{\pi}}r$, so $m(r) \to \frac{2a}{\sqrt{\pi}} = \frac{1}{\sigma\sqrt{\pi}}$.
			\item As $r \to \infty$: $\erf{ar} \to 1$, so $m(r) \sim \frac{1}{r} \to 0$.
		\end{itemize}
		
		Now establish monotonicity. The derivative is
		\begin{equation}
			m'(r) = \frac{1}{r^2}\left(\frac{2a}{\sqrt{\pi}} e^{-a^2r^2}r - \erf{ar}\right) = -\frac{a^2}{t^2}\,\phi(t),
		\end{equation}
		where $t=ar$ and $\phi(t) := \erf t - \frac{2t}{\sqrt{\pi}} e^{-t^2}$. Then
		\begin{equation}
			\phi'(t) = \frac{2}{\sqrt{\pi}} e^{-t^2} - \frac{2}{\sqrt{\pi}}\bigl(e^{-t^2} - 2t^2 e^{-t^2}\bigr)
			= \frac{4}{\sqrt{\pi}} t^2 e^{-t^2} \ge 0,
		\end{equation}
		with equality only at $t=0$.
		
		Since $\phi(0) = 0$ and $\phi'(t) \ge 0$, we have $\phi(t) > 0$ for all $t > 0$. Therefore $m'(r) < 0$ for all $r>0$, and $m(r)$ is strictly decreasing on $(0,\infty)$, attaining its supremum as $r \to 0^+$:
		\begin{equation}
			\|M\| = \lim_{r\to 0^+} m(r) = \frac{2a}{\sqrt{\pi}} = \frac{1}{\sigma\sqrt{\pi}}.
			\tag{1}
		\end{equation}
		
		\noindent\textbf{Step 3: Norm of the integral operator $S$.}
		
		We first establish a lemma.
		
		\begin{lemma}\label{lem:S-norm}
			The operator $S$ with kernel
			\begin{equation}
				S(r,r') = \frac{1}{k}\left(e^{-k|r-r'|} - e^{-k(r+r')}\right) = \frac{2}{k}\sinh(k r_<) e^{-k r_>}
			\end{equation}
			satisfies
			\begin{equation}
				\|S\|_{L^2(0,\infty)} = \frac{2}{k^2}.
			\end{equation}
		\end{lemma}
		
		\begin{proof}[Proof of Lemma]
			Let $\mathcal{L}$ be the differential operator
			\begin{equation}
				\mathcal{L} := -\frac{d^2}{dr^2} + k^2
			\end{equation}
			defined on the domain
			\begin{equation}
				D(\mathcal{L}) = \{f \in H^2(0,\infty) : f(0) = 0\}.
			\end{equation}
			
			The Green's function $G_{\mathcal{L}}$ for $\mathcal{L}$ (i.e., the solution to $(\mathcal{L} G)(r) = \delta(r-r')$ with Dirichlet boundary conditions at $r=0$ and decaying behavior at infinity) is
			\begin{equation}
				G_{\mathcal{L}}(r,r') = \frac{1}{k}\sinh(k r_<) e^{-k r_>}.
			\end{equation}
			
			Therefore the kernel of $S$ is $S(r,r')=2\,G_{\mathcal{L}}(r,r')$, i.e.\ $S = 2\mathcal{L}^{-1}$.
			
			The operator $\mathcal{L}$ is self-adjoint on $L^2(0,\infty)$ with domain $D(\mathcal{L})$. Its spectrum is purely essential:
			\begin{equation}
				\sigma(\mathcal{L}) = [k^2, \infty),
			\end{equation}
			because $\mathcal{L} = -\Delta_{1D} + k^2$ on the half-line with Dirichlet boundary, where $-\Delta_{1D}$ has spectrum $[0,\infty)$ on this domain. The constant shift $k^2$ moves the spectrum to $[k^2,\infty)$; there is no discrete spectrum because the potential is constant.
			
			Consequently,
			\begin{equation}
				\|\mathcal{L}^{-1}\| = \frac{1}{\inf \sigma(\mathcal{L})} = \frac{1}{k^2}.
			\end{equation}
			
			Hence
			\begin{equation}
				\|S\| = 2\|\mathcal{L}^{-1}\| = \frac{2}{k^2}.
				\tag{2}
			\end{equation}
		\end{proof}
		
		\noindent\textbf{Step 4: Combine the bounds.}
		
		From (1) and (2),
		\begin{equation}
			\|K_k\| \le \|S\|\,\|M\| \le \frac{1}{\sigma\sqrt{\pi}} \cdot \frac{2}{k^2} = \frac{2}{\sqrt{\pi}\,\sigma\, k^2}.
			\tag{3}
		\end{equation}
	\end{proof}
	
	\begin{corollary}[Necessary Condition for Bound States]\label{corr:lower_bound}
		If the Fredholm equation $y = K_k y$ has a nontrivial $L^2$-solution at energy $E = -k^2/2 < 0$, then
		\begin{equation}
			k^2 \le \frac{2}{\sqrt{\pi}\,\sigma}.
		\end{equation}
		Equivalently,
		\begin{equation}
			E \ge -\frac{1}{\sqrt{\pi}\,\sigma}.
		\end{equation}
	\end{corollary}
	
	\begin{proof}
		At a bound state, the operator $K_k$ has eigenvalue $1$, so
		\begin{equation}
			1 = |\lambda_{\max}(K_k)| \le \|K_k\| \le \frac{2}{\sqrt{\pi}\,\sigma\,k^2}.
		\end{equation}
		Therefore $k^2 \le \frac{2}{\sqrt{\pi}\,\sigma}$, and hence
		\begin{equation}
			E = -\frac{k^2}{2} \ge -\frac{1}{\sqrt{\pi}\,\sigma}.
		\end{equation}
	\end{proof}
	
	\begin{remark}
		The bound in Corollary~\ref{corr:lower_bound} is a necessary but not sufficient condition for the existence of a bound state. It provides a useful theoretical guarantee: for a given screening parameter $\sigma$, bound states can only exist with energies above the threshold $E_{\min} = -1/(\sqrt{\pi}\sigma)$. Numerical results in Section~\ref{sec:results} show that actual ground state energies lie significantly above this bound, indicating that the bound is conservative but rigorous.
	\end{remark}
	
	This bound is sufficient for theoretical purposes; a sharper estimate could be obtained via numerical computation of the largest singular value.
	
	\begin{proposition}[Iteration Convergence to the Ground State]\label{prop:iterconv}
		Fix $k>0$ and $\sigma>0$. Let $m(r)=\erf{r/2\sigma}/r$, let $S$ be the operator with kernel
		$S(r,r')=\frac{2}{k}\sinh(kr_<)e^{-kr_>}$, and let $K=SM$, where $M$ is multiplication by $m$ (so $K$ is the operator of Theorem~\ref{th:bond}).
		Define the symmetrised operator $\tilde K:=M^{1/2}SM^{1/2}$ on $L^2(0,\infty)$. Then:
		\begin{enumerate}
			\item[(a)] $\tilde K$ is compact, self-adjoint and non-negative. Its largest eigenvalue $\mu_0(k)=\|\tilde K\|$ is simple, with an eigenfunction $\psi_0>0$, and $\mu_0>\mu_1\ge\mu_2\ge\dots\ge0$.
			\item[(b)] $\mu_0(k)=1$ if and only if $E=-k^2/2$ is the ground-state energy $E_\ast$ of $H$ (i.e.\ $k=k_\ast$).
			\item[(c)] For $y_0(r)=re^{-kr}$ and $y_{j+1}=Ky_j$, put $\rho:=\mu_1/\mu_0<1$. Then there is a constant $C$ (depending on $y_0$, $k$, $\sigma$) with
			\[
			\left\| \frac{y_j}{\|y_j\|_{L^2}}-\frac{\phi_\ast}{\|\phi_\ast\|_{L^2}}\right\|_{L^2}\le C\,\rho^{\,j},
			\qquad \phi_\ast:=SM^{1/2}\psi_0\;(\,=\mu_0\, M^{-1/2}\psi_0>0\,),
			\]
			and $K\phi_\ast=\mu_0\phi_\ast$. For $k=k_\ast$ one has $\mu_0=1$ and $\phi_\ast$ is the ground-state reduced wave function $\phi_0$. For $k\neq k_\ast$ the iteration converges to the dominant eigenfunction of $K_k$, which is not an eigenfunction of $H$.
		\end{enumerate}
	\end{proposition}
	
	\begin{proof}
		\emph{Step 1: properties of $\tilde K$.}
		Since $S=2\mathcal L^{-1}$ with $\mathcal L=-d^2/dr^2+k^2$ (Dirichlet at $0$), $S$ is bounded, symmetric and positive. The function $m$ is positive, bounded and satisfies $m(r)\le C/(1+r)$. Hence $\tilde K$ is bounded and symmetric, and non-negative:
		$\langle u,\tilde Ku\rangle=\langle M^{1/2}u,\,S\,M^{1/2}u\rangle\ge0$.
		Its kernel is
		\[
		\kappa(r,r')=m(r)^{1/2}S(r,r')\,m(r')^{1/2}>0,
		\]
		because $\sinh(kr_<)e^{-kr_>}>0$ for $r,r'>0$. Moreover $0<S(r,r')\le\frac1k e^{-k|r-r'|}$, and
		$\int_0^\infty e^{-2k|r-r'|}(1+r')^{-1}dr'\le C'(1+r)^{-1}$, for some bounded $C'$ so
		\[
		\iint \kappa^2\,dr\,dr' \le \frac{1}{k^2}\iint m(r)m(r')e^{-2k|r-r'|}\,dr\,dr'\le C''\int_0^\infty\frac{dr}{(1+r)^2}= C''.
		\]
		Thus $\kappa\in L^2((0,\infty)^2)$, $\tilde K$ is Hilbert--Schmidt and hence compact. By the Jentzsch (Perron--Frobenius) theorem for compact operators with strictly positive kernel, the spectral radius $\mu_0=\|\tilde K\|>0$ is a simple eigenvalue with an eigenfunction $\psi_0>0$, and every other eigenvalue satisfies $|\mu_j|<\mu_0$. By compactness and non-negativity, the remaining eigenvalues are $\mu_1\ge\mu_2\ge\dots\ge0$, accumulating only at $0$. This proves (a).
		
		\emph{Step 2: relation to $H$.}
		If $u\in L^2$ satisfies $\tilde Ku=\mu u$ with $\mu\ne0$, then $\phi:=\mu^{-1}SM^{1/2}u$ satisfies $K\phi=\mu\phi$ and $M^{1/2}\phi=u$. Conversely, if $K\phi=\mu\phi$ then $u=M^{1/2}\phi$ is an eigenvector of $\tilde K$. For $\mu=1$ the equation $\phi=SM\phi=2\mathcal L^{-1}M\phi$ is equivalent to $(-\tfrac12\partial_r^2-m+\tfrac{k^2}{2})\phi=0$ with $\phi(0)=0$, i.e.\ $H\phi=E\phi$.
		Let $\phi_0>0$ be the (nodeless) ground state of $H$ at $k=k_\ast$. Then $K_{k_\ast}\phi_0=\phi_0$, so $u=M^{1/2}\phi_0$ is a positive eigenvector of $\tilde K$ with eigenvalue $1$. Since $\langle u,\psi_0\rangle>0$, $u$ is not orthogonal to $\psi_0$. Eigenvectors of a symmetric operator for different eigenvalues are orthogonal, so the eigenvalue of $u$ is $\mu_0$, and therefore $\mu_0(k_\ast)=1$. Conversely, $\mu_0(k)=1$ gives a positive eigenfunction $\phi=\mu_0^{-1}SM^{1/2}\psi_0$ of $H$, which by Sturm oscillation theory for the radial problem is the ground state. This proves (b).
		
		\emph{Step 3: convergence.}
		Put $u_j:=M^{1/2}y_j$. Then $u_{j+1}=M^{1/2}SMy_j=M^{1/2}SM^{1/2}\,u_j=\tilde Ku_j$. Moreover $u_0=M^{1/2}y_0\in L^2$, and
		\[
		c_0:=\langle u_0,\psi_0\rangle=\int_0^\infty m(r)^{1/2}\,re^{-kr}\,\psi_0(r)\,dr>0
		\]
		because the integrand is positive. Let $\{\psi_i\}$ be an orthonormal eigenbasis of $\tilde K$ and write $u_0=c_0\psi_0+w_0$, $w_0\perp\psi_0$, $r_0:=\|w_0\|$. Then
		\[
		\tilde K^{\,j}u_0=\mu_0^{\,j}\bigl(c_0\psi_0+w_j\bigr),\qquad \|w_j\|\le\rho^{\,j}\,r_0 .
		\]
		For unit-vector normalisation one has the elementary estimate
		$\bigl\|x/\|x\|-e/\|e\|\bigr\|\le 2\|x-e\|/\|e\|$, so
		\[
		\left\|\frac{u_j}{\|u_j\|}-\psi_0\right\|\le \frac{2r_0}{c_0}\,\rho^{\,j}.
		\]
		Since $y_{j+1}=SM^{1/2}u_j$, we have $$y_{j+1}/\mu_0^{j+1}=SM^{1/2}(u_j/\mu_0^j)/\mu_0\to (c_0/\mu_0)\,SM^{1/2}\psi_0=(c_0/\mu_0)\,\phi_\ast$$ 
		in $L^2$, and applying the same normalisation estimate yields the bound.
	\end{proof}
	
	\begin{remark}
		The rate is $\rho=\mu_1(k)/\mu_0(k)$. At $k=k_\ast$ this equals $\mu_1<1$, which is the "spectral gap" of $K$. It follows from the simplicity of the ground state of $H$ and Step 2, so no homotopy argument is needed. For $k\ne k_\ast$, the iterates still converge, but to the dominant eigenfunction of $K_k$, which is not an eigenfunction of $H$. This is why the iteration is combined with the variational optimisation of $k$.
	\end{remark}

	\section{Scaling Properties for Arbitrary Nuclear Charge}
	
	\subsection{Fredholm Equation with General $Z$}
	
	For a nuclear charge $Z$, the erf-screened Coulomb potential reads
	\[
	V(r) = -Z\,\frac{\erf{r/(2\sigma)}}{r}.
	\]
	The corresponding radial Schr\"odinger equation at energy $E=-k^{2}/2$ is
	transformed into the Fredholm integral equation
	\begin{equation}\label{eq:fredholm-Z}
		y(r) = -Z\int_{0}^{\infty} G_{k}(r,r')\,\frac{\erf{r'/(2\sigma)}}{r'}\,y(r')\,dr',
	\end{equation}
	where $G_{k}$ is the free‑particle Green’s function \eqref{eq:greenfree} parametrized by $k$, which is derived in Appendix~\ref{sec:green}.  
	Defining the integral operator,	which is independent of $Z$:
	\[
	K_{k}[y](r):=-\int_{0}^{\infty} G_{k}(r,r')\,\frac{\erf{r'/(2\sigma)}}{r'}\,y(r')\,dr',
	\]
	equation~\eqref{eq:fredholm-Z} becomes
	\begin{equation}\label{eq:eigen-Z}
		K_{k}y = \frac{y}{Z}.
	\end{equation}
	Thus, a bound state exists precisely when $1/Z$ is an eigenvalue of the 
	compact operator $K_{k}$.  For the pure Coulomb case ($\sigma\to0$) the 
	well‑known result $k=Z$ (in atomic units) is recovered.
	
	\subsection{Dimensional Analysis and Universal Scaling}
	
	Introduce the dimensionless variables
	\[
	\rho = Z r,\qquad \kappa = \frac{k}{Z},\qquad s = Z\sigma .
	\]
	In these variables the Hamiltonian becomes
	\[
	H = Z^{2}\!\left[-\frac{1}{2}\nabla_{\rho}^{2} 
	- \frac{\erf{\rho/(2s)}}{\rho}\right],
	\]
	and the Fredholm equation~\eqref{eq:fredholm-Z} reduces to the 
	$Z$‑independent form
	\begin{equation}\label{eq:fredholm-dimless}
		y(\rho) = - \int_{0}^{\infty} G_{\kappa}(\rho,\rho')
		\frac{\erf{\rho'/(2s)}}{\rho'}\,y(\rho')\,d\rho'.
	\end{equation}
	Consequently the whole eigenvalue problem depends only on the 
	\emph{dimensionless screening parameter} $s=Z\sigma$.  
	All physical quantities therefore exhibit the scaling
	\begin{subequations}\label{eq:scaling-laws}
		\begin{align}
			k(Z,\sigma) &= Z\,\kappa(s), \label{eq:k-scaling} \\
			E (Z,\sigma) &= Z^{2}\,\mathcal{E}(s), \label{eq:E-scaling} \\
			y_{Z}(r) &= Z^{3/2}\,y (Z r;\,s), \label{eq:y-scaling}
		\end{align}
	\end{subequations}
	where $\kappa(s)$, $\mathcal{E}(s)$ and $y(\rho;s)$ are the
	wave‑number, energy and wavefunction obtained for $Z=1$ at the
	effective screening length $\sigma=s$.

	\section{Some properties of the wavefunction gradient}\label{sec:gradient}
	Throughout this section we will assume that the function $y$ varies freely with $k$, while $\sigma$ is treated as a parameter.
	
	\begin{proposition}[Gradient]\label{prop:deriv1}
		The	gradient of the reduced wave function is given by the equation
		\begin{equation}\label{eq:prop:deriv1}
			y'  (r) = - 2\; e^{- k r} J_1(r) + 2 \cosh{( k r )} J_2(r) 
		\end{equation} 
	\end{proposition}
	\begin{proof}
		The proof follows from the observation that
		\begin{equation} 
			\sinh{(k r)}J_2^\prime(r) +e^{-k r } J_1^\prime(r) =0
		\end{equation}
		Indeed, by direct computation from eq. \eqref{eq:fredholm} we obtain
		\begin{multline}
			\frac{2}{k} e^{-k r} \frac{d}{dr} \int_0^r \frac{\erf{ \frac{r'}{2\sigma} }}{r'} \sinh{(k r')} y(r') \, dr' \\
			+ \frac{2}{k} \sinh{(k r)} \frac{d}{dr} \int_r^\infty \frac{\erf{ \frac{r'}{2\sigma} }}{r'} e^{-k r'} y(r') \, dr' \\
			= \frac{2}{k} e^{-k r} \frac{\erf{ \frac{r}{2\sigma} }}{r} \sinh(k r) y(r) +
			\frac{2}{k} \sinh{(k r)}\left(- \frac{\erf{ \frac{r}{2\sigma} }}{r} e^{-k r} y(r) \right) =0
		\end{multline}
		identically. Therefore, to compute the derivative we only need to differentiate the pre-factors of $J_1$ and $J_2$.
	\end{proof}

	\subsection{Gradient bounds at the origin}\label{sec:gradient_bounds}
	
	Throughout this section, we assume that the reduced wave function \(y\) varies freely with \(k\), while the screening parameter \(\sigma\) is held fixed.
		
	To compute the bounds of the gradient we need to state first a technical lemma.
	\begin{lemma}\label{th:J3}
		Let $a=1/(2\sigma)$ and let $y\ge0$ be bounded on $[0,\infty)$ with $\|y\|_\infty<\infty$. Then
		\[
		0\le J_2(r)\le J_2(0)\le \|y\|_\infty\,J_3,
		\qquad
		J_3:=\int_0^\infty\frac{\operatorname{erf}(at)}{t}\,e^{-kt}\,\mathrm{d}t ,
		\]
		with the strict inequality $J_2(0)<\|y\|_\infty J_3$ unless $y\equiv\|y\|_\infty$ almost everywhere on $(0,\infty)$, 
		\begin{equation}\label{eq:J3}
			J_3=\frac{k}{\sqrt{\pi}\,a}\,{}_2F_2\!\left(\tfrac12,1;\tfrac32,\tfrac32;\tfrac{k^2}{4a^2}\right)
			-\tfrac12\operatorname{Ei}\!\left(\tfrac{k^2}{4a^2}\right).
		\end{equation}
		and ${}_2F_2 (a,b; c,d; z)$ denotes the generalized hypergeometric function while $\operatorname{Ei}$ denotes the exponential integral.  
	\end{lemma}
	
	\begin{proof}[Sketch]
		The proof is obtained by Mellin transform/Parseval identity applied to $J_3$.
		A detailed proof is given in \ref{sec:proofL1}.
	\end{proof}

	Therefore, we claim the following bound:

	\begin{proposition}[Gradient of the ground state at the origin]\label{prop:gradbound_exact}
		Let $y$ be the $L^2$-normalized reduced ground state of $H$, $E_\ast=-k_\ast^2/2$. Then
		\[
		y'(0)=2\int_0^\infty e^{-k_\ast r}m(r)\,y(r)\,dr\ \le\ 2\|y\|_\infty J_3(k_\ast,\sigma),
		\qquad
		y'(0)\le 2\sqrt{\frac{J_3(2k_\ast,\sigma)}{\sigma\sqrt\pi}} .
		\]
		For $y_0=re^{-kr}$ and $y_1=K_ky_0$ (unnormalized),
		\[
		y_1'(0)=\frac1k\,e^{x^2}\erfc{x}\ \le\ \frac1k,\qquad x=2k\sigma .
		\]
		Moreover, at $k=k_\ast$ and with $\hat y_j:=y_j/\|y_j\|$,
		\[
		\bigl|\hat y_{j+1}'(0)-y'(0)\bigr|\ \le\ C'\rho^{\,j},
		\]
		where $C'$ depends on $C$ of Theorem~\ref{th:conv}, on $\|e^{-k_\ast\cdot}m\|_{L^2}$ and on the ratio $\|y_{j+1}\|/\|y_j\|$ (which tends to $1$).
	\end{proposition}

	\begin{proof}
		Throughout, $k=k_\ast$, $m(r)=\erf{r/2\sigma}/r$, $\|m\|_\infty=1/(\sigma\sqrt\pi)$ (eq.~(1)), and
		$J_3(k,\sigma)=\int_0^\infty m(r)e^{-kr}\,dr$.
		
		\emph{Step 1: the identity for $y'(0)$.}
		The ground state satisfies $y=\int G_kVy$, i.e.
		$y=\tfrac2k e^{-kr}J_1+\tfrac2k\sinh(kr)J_2$ with $J_1,J_2$ built from $y$ as in
		eq.~\eqref{eq:fredholm}. By Prop.~\ref{prop:deriv1},
		$y'(r)=-2e^{-kr}J_1(r)+2\cosh(kr)J_2(r)$. Since $J_1(0)=0$,
		\begin{equation}\label{eq:yp0id}
			y'(0)=2J_2(0)=2\int_0^\infty e^{-kr'}\,m(r')\,y(r')\,dr' .
		\end{equation}
		The ground state is nodeless, so $y>0$ and $y'(0)>0$. Also $y\in H^1_0$ gives
		$y(r)^2\le 2\|y\|_2\|y'\|_2<\infty$, so $\|y\|_\infty<\infty$.
		
		\emph{Step 2: the two bounds.}
		Since $0<y\le\|y\|_\infty$, \eqref{eq:yp0id} gives
		\[
		y'(0)\le 2\|y\|_\infty\int_0^\infty m\,e^{-kr}\,dr=2\|y\|_\infty J_3(k,\sigma).
		\]
		By Cauchy--Schwarz inequality applied to $(m^{1/2}e^{-kr})(m^{1/2}y)$,
		\begin{multline*}
		y'(0)\le 2\Bigl(\int_0^\infty m\,e^{-2kr}dr\Bigr)^{1/2}\Bigl(\int_0^\infty m\,y^2dr\Bigr)^{1/2}
		\le 2\sqrt{J_3(2k,\sigma)}\,\sqrt{\|m\|_\infty\|y\|_2^2}
		= \\
		 2\sqrt{\frac{J_3(2k,\sigma)}{\sigma\sqrt\pi}} .
		\end{multline*}
		
		\emph{Step 3: the first iterate.}
		For $y_0=re^{-kr}$ one has $m\,y_0=\erf{r/2\sigma}e^{-kr}$, and the same computation as in
		Step~1 (with $y$ replaced by $y_0$ inside $J_2$) gives
		\[
		y_1'(0)=2\int_0^\infty \erf{\tfrac r{2\sigma}}\,e^{-2kr}\,dr .
		\]
		For $a>0,\ p>0$, integration by parts gives
		$$
		\int_0^\infty\erf{ar}e^{-pr}dr=\frac1p\int_0^\infty\frac{2a}{\sqrt\pi}e^{-a^2r^2-pr}dr
		=\frac1p\,e^{p^2/4a^2}\erfc{p/2a}.
		$$ 
		With $p=2k$, $a=1/2\sigma$, $x=2k\sigma$,
		\[
		y_1'(0)=\frac1k\,e^{x^2}\erfc{x}\ \le\ \frac1k,
		\]
		because $\frac{d}{dx}\bigl(e^{x^2}\erfc{x}\bigr)=2xe^{x^2}\erfc{x}-\frac2{\sqrt\pi}<0$ for $x\ge0$
		(using $e^{x^2}\erfc{x}<1/(x\sqrt\pi)$) and the value at $x=0$ is $1$.
		
		\emph{Step 4: convergence of the normalized derivative.}
		Define the bounded linear functional $\ell(f):=2\int_0^\infty e^{-kr}m(r)f(r)\,dr$ on $L^2$. Then
		\begin{equation}\label{eq:ellbound}
			|\ell(f)|\le L\|f\|_2,\qquad
			L:=2\|e^{-k\cdot}m\|_{L^2}\le \frac{\sqrt2}{\sigma\sqrt{\pi k}} .
		\end{equation}
		By the argument of Step~1 applied to $y_{j+1}=Ky_j$, $y_{j+1}'(0)=\ell(y_j)$; moreover
		$K\hat\phi_0=\hat\phi_0$ with $\hat\phi_0:=\phi_0/\|\phi_0\|=y$, so $y'(0)=\ell(\hat\phi_0)$.
		Put $\hat y_j=y_j/\|y_j\|$ and $T_j:=\|K\hat y_j\|=\|y_{j+1}\|/\|y_j\|>0$. Then
		$\hat y_{j+1}'(0)=\ell(\hat y_j)/T_j$, and
		\[
		\hat y_{j+1}'(0)-y'(0)=\frac{\ell(\hat y_j)-\ell(\hat\phi_0)}{T_j}+\ell(\hat\phi_0)\Bigl(\frac1{T_j}-1\Bigr).
		\]
		By Theorem~\ref{th:conv}, $\|\hat y_j-\hat\phi_0\|\le C\rho^{\,j}$. Hence
		$|\ell(\hat y_j)-\ell(\hat\phi_0)|\le LC\rho^{\,j}$ by \eqref{eq:ellbound}. Since $\|K\hat\phi_0\|=1$ and
		$\|K\|\le \frac{2}{\sqrt\pi\sigma k^2}$ (Theorem~\ref{th:bond}),
		\[
		|T_j-1|=\bigl|\|K\hat y_j\|-\|K\hat\phi_0\|\bigr|\le\|K\|C\rho^{\,j}.
		\]
		Let $j_0$ be such that $\|K\|C\rho^{\,j_0}\le\tfrac12$. For $j\ge j_0$ we have $T_j\ge\tfrac12$ and
		\[
		|\hat y_{j+1}'(0)-y'(0)|\le 2LC\rho^{\,j}+L\cdot 2\|K\|C\rho^{\,j}
		=2LC\,(1+\|K\|)\,\rho^{\,j}=:C'\rho^{\,j},
		\]
		using $|\ell(\hat\phi_0)|=y'(0)\le L$. For the finitely many $j<j_0$, the left side is finite, since $T_j>0$,
		so enlarging $C'$ gives the bound for all $j\ge0$.
	\end{proof}
	

	\section{Approximate Analytical Solution for First Iteration}
	
	\begin{theorem}[Analytical First Order Approximation]
		For the screened potential given by eq. \eqref{eq:erfcoul} and trial function $y_0= r e^{-kr}$
		the first order Fredholm integral kernels admit the closed-form representation
		\begin{equation}\label{eq:y1}
			y_1(r) = \frac{2}{k} e^{-k r} I_1(r) + \frac{2}{k} \sinh(k r) I_2(r),
		\end{equation}
		where:
		\begin{multline}
			I_1(r)   = \left( \frac{r}{2}  + \frac{e^{-2k r} }{4k} \right)  \erf{ \frac{r}{2\sigma} }+ \frac{\sigma}{\sqrt{\pi}} \left( e^{-r^2/(4\sigma^2)} - 1 \right)  \\
			- \frac{1}{4k} e^{4k^2\sigma^2} \left[ \erf{ \frac{r}{2\sigma} + 2k\sigma} - \erf{ 2k\sigma} \right] \label{eq:I1}  
		\end{multline}
		and
		\begin{equation}
			I_2(r) = \frac{1}{2k} e^{4k^2\sigma^2}   \erfc{ \frac{r}{2\sigma} + 2k\sigma}  + \frac{1}{2k} e^{-2k r} \erf{ \frac{r}{2\sigma} }. \label{eq:I2}
		\end{equation}
		
	\end{theorem}
	\begin{proof}[Sketch]
		The derivation follows from integration by parts and completion of squares in Gaussian integrals. For $I_1(r)$ and $I_2(r)$, we use $\int \exp(-ax)\text{erf}(bx)dx$ formulas from Gradshteyn and Ryzhik \cite{gradshteyn2007}. Full details are provided in \ref{sec:iter}.
	\end{proof}
	
	\begin{corollary}
		The	first derivative of $y_1(r)$ is given by
		\begin{equation}\label{eq:derivy1}
			y'_1 (r) = - 2\; e^{- k r} I_1(r) + 2 \cosh{( k r )} I_2(r) 
		\end{equation} 
	\end{corollary}
	\begin{proof}
		The proof follows from the application of the chain rule followed by the result of Prop.~\ref{prop:deriv1}.
	\end{proof}
	\subsection{Variational Energy Functionals}
	An estimate for the eigen energy of the Sturm–Liouville problem
	can be obtained by minimization of the energy functional (i.e. the so-called Rayleigh quotient) defined as
	\begin{equation}
		E_m[k] := \frac{\langle y_m | \hat{H} | y_m \rangle}{\langle y_m | y_m \rangle} = \frac{T_m[k] + V_m[k]}{N_m[k]},
		\label{eq:rayleigh}
	\end{equation}
	where:
	\begin{align}
		N_m[k] &:= \int_0^\infty |y_m(r;k)|^2 \, dr, \\
		T_m[k] &:= \frac{1}{2}\int_0^\infty |y_m'(r;k)|^2 \, dr, \\
		V_m[k] &:= -\int_0^\infty \frac{\erf{r/2\sigma} }{r}|y_m(r;k)|^2 \, dr.
	\end{align}
	are the normalization, kinetic and potential energy terms, respectively.
	The initial estimate can be obtained in closed form. We start from $y_0(r;k)$ and obtain
	\begin{proposition}[Zeroth-order Energy function]
		\begin{equation}\label{eq:E0}
			E_0[k]= k\left( 8 {{k}^{2}} {{\sigma}^{2}}- 1\right)  {{ e}^{4 {{k}^{2}} {{\sigma}^{2}}}} \erfc{ 2 k \sigma} - \frac{4 {{k}^{2}} \sigma}{\sqrt{\pi }}+\frac{{{k}^{2}}}{2}
		\end{equation}
		where $T= k^2/2$ is the kinetic term and the rest is the potential term.
	\end{proposition}
	\begin{proof}
		The proof follows by direct computation. 
		First, compute the normalization
		\begin{equation}\label{eq:N0}
			N_0 = \int_{0}^{\infty}  r^2 e^{- 2 k r} dr = \frac{1}{4 k^3}
		\end{equation}
		Further,   the kinetic energy term is
		\begin{equation}\label{eq:T0}
			T_0 = \frac{1}{2} \int_{0}^{\infty} (k r -1)^2 e^{- 2 k r} dr = \frac{1}{8 k}  
		\end{equation}
		To compute the potential term consider the integral dependence of the parameter \textit{a}
		\[
		P_1 := -  \int r e^{- 2 k r} \erf{a r} dk = \frac{1}{2} e^{- 2 k r} \erf{a r}
		\]
		up to a constant. Then set the constant to zero and compute the improper integral
		\[
		P_2 (a) := \int_{0}^{\infty} P_1 dr = \frac{1}{4 k} e^{k^2/a^2} \erfc{\frac{k}{a}}
		\]
		from the Laplace transform pair of $\erf{ar}$
		and differentiate in  \textit{a}
		\[
		\frac{d}{da} P_2 (a)   = \frac{1}{4 k^2 a^2}( 2 k^2 -a^2) e^{k^2/a^2}\erfc{\frac{k}{a}} - \frac{1}{2 \sqrt \pi a k} 
		\]
		Finally, substitute to obtain
		\begin{equation}\label{eq:V0}
			V_0= - \frac{d}{da} P_2 \Bigg|_{a=1/(2 \sigma)} = -	\frac{1}{\sqrt{\pi}\sigma k} + \frac{\sigma^2}{k^2}\left(2 k^2 - \frac{1}{4 \sigma^2} \right)e^{4k^2 \sigma^2} \erfc{2k \sigma} 
		\end{equation}
		Therefore, it follows that
		\begin{align}
			\langle T \rangle_0 &= \frac{T_0}{N_0}= \frac{k^{2}}{2}, \\ 
			\langle V \rangle_0 &= \frac{V_0}{N_0}= 
			k\bigl(8k^{2}\sigma^{2}-1\bigr)e^{4k^{2}\sigma^{2}}\erfc{2k\sigma}
			-\frac{4k^{2}\sigma}{\sqrt{\pi}}
		\end{align}
	\end{proof}
	\begin{remark}
		For numerical stability, the equation can be rewritten in terms of the Faddeeva function $w(z):=\exp{(-z^2)} \erfc{-iz} $ as
		\begin{equation}\label{eq:E0fadd}
			E_0[k] =   k \left( 2 x^2-1  \right) \Re [w(i x)] -  	\frac{2 k x}{\sqrt{\pi } }   + \frac{k^2}{2}, \quad x = 2 k \sigma 
		\end{equation}
	\end{remark}
	
	\begin{corollary}
		For $\sigma=0$ the energy reduces to the energy of the Coulomb system:
		\[
		E_0 = -k + \frac{k^2}{2} \Longrightarrow E_0'= -1 +k \Longrightarrow E_0''=1
		\]
		Hence, the  minimum is at $k=1$ where $E_0=-\tfrac12$.
	\end{corollary}
	\begin{proposition}[Energy bounds for the zeroth-order functional]\label{prop:E0bounds}
		Let $\sigma>0$. Then
		\begin{enumerate}
			\item[(i)] $E_0[k]>\tfrac{k^2}{2}-k\ \ge\ -\tfrac12$ for every $k>0$;
			\item[(ii)] $E_0[\,\cdot\,]$ attains its minimum at some $k^*\in(0,\infty)$, and
			$-\tfrac12<E_0[k^*]<0$;
			\item[(iii)] $E_0'[k]>0$ for all $k\ge 1$; consequently every minimizer satisfies $k^*<1$.
		\end{enumerate}
	\end{proposition}
	
	\begin{proof}
		Write $\hat y_k=y_0/\|y_0\|$ with $y_0=re^{-kr}$, so that
		$\hat y_k(r)^2=4k^3r^2e^{-2kr}$ and
		$E_0[k]=\tfrac{k^2}{2}+W(k)$, where
		\[
		W(k):=-\int_0^\infty \frac{\erf{r/2\sigma}}{r}\,\hat y_k(r)^2\,dr .
		\]
		Substituting $u=kr$ gives
		\begin{equation}\label{eq:Wg}
			W(k)=-k\,g(k),\qquad 
			g(k):=\int_0^\infty 4u\,e^{-2u}\,\erf{\frac{u}{2k\sigma}}\,du .
		\end{equation}
		Since the integral $\int_0^\infty 4ue^{-2u}\,du=1$ and $0<\erf{t}<1$ for $t>0$, it follows that 
		\begin{equation}\label{eq:g01}
			0<g(k)<1 \quad\text{for all }k>0 .
		\end{equation}
		Moreover, $\erf{u/(2k\sigma)}$ is strictly decreasing in $k$ for  $u>0$, so $g$ is strictly decreasing, i.e.\ $g'(k)<0$. Differentiation under the integral sign is justified by the  Dominated Convergence theorem, since $\partial_k\erf{u/(2k\sigma)}=-\frac{u}{\sqrt\pi\,k^2\sigma}\,e^{-u^2/(4k^2\sigma^2)}$ is bounded by an integrable multiple of $u$ on compact $k$-sets.
		
		\emph{(i)} By \eqref{eq:Wg} and \eqref{eq:g01}, $W(k)>-k$, hence
		$E_0[k]>\tfrac{k^2}{2}-k=\tfrac12(k-1)^2-\tfrac12\ge-\tfrac12$.
		(This is the statement that $V_\sigma>V_{\rm Coul}$ pointwise.)
		
		\emph{(ii)} $E_0$ is continuous on $I=(0,\infty)$. 
		As $k\to0^+$, $g(k)\to1$ (since $\erf{u/2k\sigma}\to1$) and $E_0[k]=-k+\mathcal{O}(k^2)$, so $E_0<0$ for small $k$ and $E_0'(0^+)=-1$. As $k\to\infty$, $E_0[k]\ge \tfrac{k^2}{2}-k\to+\infty$. Hence $\inf_{k>0}E_0<0$ is attained at some $k^*\in(0,\infty)$, with $E_0[k^*]<0$. By (i), $E_0[k^*]>-\tfrac12$.
		
		\emph{(iii)} From \eqref{eq:Wg},
		\[
		E_0'[k]=k-g(k)-k\,g'(k)=k-g(k)+k\,|g'(k)| .
		\]
		For $k\ge1$, using \eqref{eq:g01} and $g'<0$,
		$E_0'[k]> k-1\ge 0$ (the term $k|g'(k)|$ is positive and $g<1$).
		Thus $E_0$ is strictly increasing on $[1,\infty)$, so no minimizer can lie in $[1,\infty)$, i.e.\ $k^*<1$.
	\end{proof}
	
	\begin{corollary}[Upper bound of the wavenumber]\label{corr:k}
		For $\sigma>0$ the minimizer of $E_0[k]$ satisfies $k^*<1$, and $E_0[k^*]\in(-\tfrac12,0)$.
	\end{corollary}

	\begin{remark}
		The following form in terms of the Faddeeva function the energy gradient 
		\begin{equation}\label{eq:dEdk}
			{E_0'(k) = \left(4x^4 + 4x^2 - 1\right) \Re [w\! \left(ix\right)] - \frac{2x(2x^2 + 1)}{\sqrt{\pi}} + k}
		\end{equation} 
		where  $x = 2k\sigma$, can be beneficial for numerical implementation.
	\end{remark}
	\begin{lemma}[Lyapunov property of Energy function]\label{lem:lyapunov}
		For each $\sigma>0$, $E_0[k]$ is strictly convex on $I=(0,\infty)$, with $E_0''[k]>1$.
		Hence $E_0$ has a unique minimizer $k^*(\sigma)$, which is the unique zero of
		$E_0'$; $E_0$ decreases on $(0,k^*)$ and increases on $(k^*,\infty)$. Moreover,
		along the gradient flow $\dot k=-\eta E_0'[k]$ the shifted functional
		$V[k]:=E_0[k]-E_0[k^*]\ge0$ is a Lyapunov function, $\dot V=-\eta (E_0'[k])^2\le0$,
		and it decays as $V(t)\le V(0)e^{-2\eta t}$.
	\end{lemma}
	
	\begin{proof}
		By dilation, $E_0[k]=\tfrac{k^2}{2}-h(k)$ with
		$h(k)=\int_0^\infty 4u\,e^{-2u}\,\varphi_u(k)\,du$ and
		$\varphi_u(k)=k\,\erf{c/k}$, $c=u/(2\sigma)$. A direct computation gives
		$\varphi_u''(k)=-\frac{4c^3}{\sqrt\pi\,k^4}e^{-c^2/k^2}<0$, so $h''<0$ and
		$E_0''=1-h''>1$. Since $E_0'(0^+)=-1$ and $E_0'(k)\to\infty$, the increasing
		function $E_0'$ has exactly one zero $k^*$. Finally $\dot V=E_0'\dot k=-\eta(E_0')^2$,
		and strong convexity gives $(E_0')^2\ge 2V$, so $\dot V\le-2\eta V$.
	\end{proof}
	
\begin{proposition}[Variational upper bound]\label{prop:varbound}
	Let $E_\ast$ be the ground-state energy of $H$. For every $m\ge0$ and $k>0$,
	\[
	E_m[k]\ \ge\ E_\ast, \qquad\text{so}\qquad \min_{k>0}E_m[k]\ \ge\ E_\ast ,
	\]
	with equality iff $y_m(\cdot;k)$ is proportional to the ground state $\phi_0$.
\end{proposition}
This is the Rayleigh--Ritz principle~\cite{griffiths2018,landau1977}; it applies
since $y_m\in H^1_0(0,\infty)$ and $\int m\,y_m^2<\infty$.
	
	%
	\subsection{Scaling Implications for the Variational Calculation}
	For the zeroth‑order trial function $y_{0}(r)=r e^{-kr}$ the energy
	functional~\eqref{eq:E0} generalises to
	\begin{equation}\label{eq:E0-Z}
		E_{0}[k;Z,\sigma] = \frac{k^{2}}{2} 
		+ Zk\bigl(8k^{2}\sigma^{2}-1\bigr)e^{4k^{2}\sigma^{2}}\erfc{2k\sigma}
		- \frac{4Zk^{2}\sigma}{\sqrt{\pi}} .
	\end{equation}
	Inserting $k=Z\kappa$ and $\sigma=s/Z$ gives
	\[
	E_{0}[k;Z,\sigma] = Z^{2}
	\Bigl[\frac{\kappa^{2}}{2} 
	+ \kappa\bigl(8\kappa^{2}s^{2}-1\bigr)e^{4\kappa^{2}s^{2}}\erfc{2\kappa s}
	- \frac{4\kappa^{2}s}{\sqrt{\pi}}\Bigr],
	\]
	confirming the scaling~\eqref{eq:E-scaling}.  Minimisation with respect
	to $\kappa$ yields a universal function $\kappa_{\mathrm{opt}}(s)$; 
	the optimal wave number for any $(Z,\sigma)$ is then obtained from
	\eqref{eq:k-scaling}.  The same scaling holds for the Fredholm iterates
	$y_{1},y_{2},\dots$ because the integral kernel in~\eqref{eq:fredholm-dimless}
	is $Z$‑independent.

	\subsection{The Virial Theorem Relation under Scaling}
	
	The identity $\Delta_{0}[k]=k\,dE_{0}/dk$ proved in 
	Proposition~\ref{prop:vireq} remains valid for arbitrary $Z$.  In dimensionless
	variables it reads
	\[
	\Delta_{0}[k;Z,\sigma] = Z^{2}\,
	\Bigl[\kappa\frac{d\widetilde{E}_{0}}{d\kappa}\Bigr],
	\qquad 
	\widetilde{E}_{0}(\kappa;s)=\frac{E_{0}[Z\kappa;Z,s/Z]}{Z^{2}} .
	\]
	Hence the proportionality between the virial deviation and the energy
	derivative is preserved by the scaling, and energy stationarity 
	($dE_{0}/dk=0$) continues to imply  satisfaction of the virial
	theorem for the one‑parameter exponential ansatz, irrespective of
	the nuclear charge.
	
	The scaling relations~\eqref{eq:scaling-laws} reduce the two‑parameter
	space $(Z,\sigma)$ to a single lumped parameter $s=Z\sigma$.  
	This observation could simplify systematic studies of
	erf‑screened systems across different atoms, molecules, or
	materials where the effective screening length $\sigma$ may vary
	with the dielectric environment while the nuclear charge $Z$ is
	fixed by the chemical species.
	
	In the  HSE family of density functionals,
	the screening parameter $\omega=0.11$~bohr$^{-1}$ corresponds to
	$\sigma \approx 4.55$~bohr.  For a hydrogen‑like centre ($Z=1$)
	this gives $s\approx 9.1$; for a heavier atom with $Z>1$ the
	effective dimensionless screening $s=Z\sigma$ becomes larger,
	weakening the binding accordingly.  The scaling laws derived here
	allow the results obtained for $Z=1$ to be immediately transferred
	to intermediate $Z$, where relativistic effects are negligible, providing a unified description of erf‑screened
	Coulomb problems in the context of range‑separated hybrid
	functionals.
	
	\section{Virial Theorem for the erf-Screened Coulomb Potential}
	
	For a stationary state of the Hamiltonian 
	\( H = T + V(r) \) with the erf-screened Coulomb potential, 
	the quantum Virial theorem reads
	\[
	2\langle T \rangle = \Big\langle r \frac{dV}{dr} \Big\rangle .
	\]
	A direct calculation gives
	\[
	r\frac{dV}{dr} = -V(r) - \frac{1}{\sigma\sqrt{\pi}} \, e^{-r^{2}/(4\sigma^{2})},
	\]
	so that
	\begin{equation}\label{eq:virial-exact}
		2\langle T \rangle + \langle V \rangle
		= -\frac{1}{\sigma\sqrt{\pi}} \,
		\big\langle e^{-r^{2}/(4\sigma^{2})} \big\rangle .
	\end{equation}
	This relation is exact for any eigenstate of \(H\).
	For a trial, i.e. an approximate eigenfunction, we define the Virial Deviation functional as 
	\begin{equation}
		\Delta_0[k]:= 2\langle T\rangle_0+\langle V\rangle_0 + \frac{1}{\sigma\sqrt{\pi}} \,
		\big\langle e^{-r^{2}/(4\sigma^{2})} \big\rangle_0
	\end{equation} 
	For the trial function \(y_{0}(r)=r e^{-kr}\) we obtain the expectation
	\begin{equation}\label{eq:expgauss}
		\big\langle e^{-r^{2}/(4\sigma^{2})} \big\rangle =
		8\sqrt{\pi}\,k^{3}\sigma^{3}
		\Bigl[ e^{4k^{2}\sigma^{2}}\erfc{2k\sigma}\,
		\bigl(1+8k^{2}\sigma^{2}\bigr)
		-\frac{4k\sigma}{\sqrt{\pi}} \Bigr],
	\end{equation}
	so that after some algebra one obtains
	\begin{equation}
		\begin{aligned}
			\Delta_{0}[k] &=
			k^{2}
			-\frac{4k^{2}\sigma}{\sqrt{\pi}}
			-\frac{32k^{4}\sigma^{3}}{\sqrt{\pi}} \\ 
			&\quad+
			k\,e^{4k^{2}\sigma^{2}}\erfc{2k\sigma}\,
			\bigl(64k^{4}\sigma^{4}+16k^{2}\sigma^{2}-1\bigr).
		\end{aligned}
	\end{equation}
	or in terms of the Faddeeva function
	\begin{equation}
		\Delta_{0}[k] = k^2 - \frac{2 k x (1+ 2x^2)}{\sqrt{\pi}} + k (4x^4 +4 x^2 -1) \Re  \left[ w\! \left(i x \right)\right] 
	\end{equation}
	For the Coulomb limit (\(\sigma\to0\), i.e. $ x \to 0$) with the exact ground‑state
	parameter \(k=1\) one finds \(\Delta_{0}\to 0\).  
	When the parameter \(k\) is chosen by minimizing the variational energy
	\(E_{0}[k]\), the magnitude \(|\Delta_{0}|\) quantifies how
	closely the trial function satisfies the exact virial relation; it thus
	provides an additional quality criterion beyond the energy itself.

	\begin{proposition}[Virial Deviation Equivalence for $y_0$]\label{prop:vireq}
		For the trial function $y_0(r;k) = r e^{-kr}$ with $k>0$, the virial deviation 
		$\Delta_0[k] \equiv 2\langle T\rangle + \langle V\rangle + 
		\frac{1}{\sigma\sqrt{\pi}}\langle e^{-r^2/(4\sigma^2)}\rangle$ 
		and the derivative of the variational energy $E_0[k]$ satisfy the exact relation
		\begin{equation}\label{eq:virial-energy-relation}
			\Delta_0[k] = k  \frac{dE_0}{dk},
		\end{equation}
		where $E_0[k]$ is given by eq. \eqref{eq:E0}.
		Consequently, 
		\[
		\frac{dE_0}{dk} = 0 \quad \Longleftrightarrow \quad \Delta_0[k] = 0
		\] for any $k>0$.
		That is, energy stationarity is equivalent to satisfaction of the virial theorem 
		(in expectation) for the single-parameter exponential ansatz.
	\end{proposition}
	
	\begin{proof}
		We compute both quantities explicitly from the above definitions.
		
		For $y_0(r)=re^{-kr}$, the expectation values are
		\begin{align*}
			\langle T\rangle &= \frac{k^2}{2}, \\
			\langle V\rangle &= k\left(8k^2\sigma^2 - 1\right)e^{4k^2\sigma^2}
			\erfc{2k\sigma} - \frac{4k^2\sigma}{\sqrt{\pi}}, \\
			\big\langle e^{-r^2/(4\sigma^2)}\big\rangle &= 
			8\sqrt{\pi}k^3\sigma^3\Bigl[e^{4k^2\sigma^2}\erfc{2k\sigma}
			\bigl(1+8k^2\sigma^2\bigr) - \frac{4k\sigma}{\sqrt{\pi}}\Bigr].
		\end{align*}
		Substituting into $\Delta_0[k] = 2\langle T\rangle + \langle V\rangle 
		+ \frac{1}{\sigma\sqrt{\pi}}\langle e^{-r^2/(4\sigma^2)}\rangle$ gives
		\begin{equation}\label{eq:Delta-explicit}
			\begin{aligned}
				\Delta_0[k] = &\ k\bigl(64k^4\sigma^4 + 16k^2\sigma^2 - 1\bigr)
				e^{4k^2\sigma^2}\erfc{2k\sigma} \\
				&+ k\Bigl[k - \frac{4k\sigma}{\sqrt{\pi}}\bigl(1+8k^2\sigma^2\bigr)\Bigr].
			\end{aligned}
		\end{equation}
		after algebraic simplification.
		
		Multiplying eq. \eqref{eq:dEdk} by $k$ gives
		\[
		k\frac{dE_0}{dk} = k\bigl(64k^4\sigma^4 + 16k^2\sigma^2 - 1\bigr)
		e^{4k^2\sigma^2}\erfc{2k\sigma}
		+ k^2 - \frac{4k^2\sigma}{\sqrt{\pi}}\bigl(8k^2\sigma^2+1\bigr),
		\]
		which is identical to \eqref{eq:Delta-explicit}. Hence
		\[
		\Delta_0[k] = k  \frac{dE_0}{dk}.
		\]
		For $k>0$, the factor $k$ is non‑zero, so $\Delta_0[k]=0$ if and only if 
		$dE_0/dk=0$.
	\end{proof}
	\begin{remark}
		The factor $k$ in \eqref{eq:virial-energy-relation} originates from the 
		scaling properties of $y_0(r;k)=r e^{-kr}$. Writing $y_0(r;k)=k^{3/2}f(kr)$ 
		with $f(u)=u e^{-u}$, variation of $k$ corresponds to a dilation of the radial coordinate. The virial theorem for dilation‑generated variations is precisely 
		the condition $dE/dk=0$; the extra factor $k$ is the Jacobian of the transformation $r \mapsto k r$.
	\end{remark}
	\begin{corollary}
		The optimal variational parameter $k_0^* = \arg\min\limits_{k>0} E_0[k]$ satisfies the 
		Virial theorem exactly: $\Delta_0[k_0^*]=0$.
	\end{corollary}
	\begin{proof}
		At the minimum, $dE_0/dk\big|_{k=k_0^*}=0$. By Prop.~\ref{prop:vireq}, 
		$\Delta_0[k_0^*] = k_0^* \cdot 0 = 0$.
	\end{proof}

	\subsection{Implications for the Fredholm Iterates}
	
	For the first Fredholm iterate $y_1(r;k)$, the simple scaling form is lost 
	because the kernel integrals $I_1(r)$ and $I_2(r)$ introduce additional 
	$\operatorname{erfc}$ functions. Consequently, 
	the relation $\Delta_1[k] = k\, dE_1/dk$ no longer holds exactly. 
	Nevertheless, numerical results in Table~\ref{tab:virial-y1} show that 
	the energy‑optimal and virial‑optimal values of $k$ for $y_1$ differ by less than $1\%$, indicating that the Fredholm iteration preserves the link between energy stationarity and virial condition satisfaction approximately.

	\section{Homotopy analysis}\label{sec:homotopy}
	
	\subsection{Matched Harmonic Oscillator (HO) Homotopy}\label{sec:harmonic-oscillator-homotopy}
	
	Let $V(r)=-\erf{\frac{r}{2\sigma}}/r$ and let $V_{\rm ho}$ be the quadratic Taylor
	approximation of $V$ at $r=0$,
	\begin{equation}\label{eq:Vho}
		V_{\rm ho}(r)=V_0+\tfrac12\kappa r^2,\qquad
		V_0=-\frac{1}{\sqrt{\pi}\,\sigma},\quad \kappa=\frac{1}{6\sqrt{\pi}\,\sigma^3},
	\end{equation}
	which follows from
	\[
	V(r) = -\frac{1}{\sqrt{\pi}\sigma} + \frac{r^2}{12\sqrt{\pi}\sigma^3} + O\!\left(\frac{r^4}{\sigma^5}\right).
	\]
	Here $\kappa$ is the spring constant and should not be confused with the decay parameter $k$.
	We define a homotopic Hamiltonian connecting the harmonic oscillator to the target
	erf-screened system \cite{he1999},
	\begin{equation}\label{eq:homhamail}
		H(\lambda) := -\frac{1}{2} \frac{d^2}{dr^2} + (1-\lambda) V_{\text{ho}}(r) + \lambda V(r),
		\qquad \lambda\in[0,1].
	\end{equation}
	For $\lambda=0$ we recover the harmonic oscillator $H(0)=H_{\rm ho}$; for $\lambda=1$ we obtain
	the full erf-screened Hamiltonian $H(1)$. We work with the radial function $y=\sqrt{4\pi}\,r\psi$,
	$y(0)=0$; the 3D $s$-state and its radial reduction have the same energy.
	For $\lambda<1$ the operator $H(\lambda)$ has compact resolvent, and the continuity of
	its eigenvalues follows from Kato's perturbation theory \cite{kato1995}; the endpoint
	$\lambda=1$, where $H(1)$ has essential spectrum $[0,\infty)$, is treated in
	Lemma~\ref{lem:homomono}.
	
	We compute the first-order energy correction $\Delta E=\frac{dE}{d\lambda}\big|_{\lambda=0}$
	to estimate the ground-state energy. The Hellmann--Feynman theorem \cite{Messiah1999} gives
	\begin{equation}
		\frac{dE(\lambda)}{d\lambda} = \frac{\Braket{y(\lambda) | \frac{\partial H}{\partial \lambda} | y(\lambda)}}{\Braket{y(\lambda) | y(\lambda)}},
		\qquad
		\frac{\partial H}{\partial \lambda} = V(r)-V_{\text{ho}}(r),
	\end{equation}
	as follows from eq.~\eqref{eq:homhamail}.
	
	\subsection{Ground State Solution}\label{sec:ground}
	
	The 3D isotropic harmonic oscillator has ground state wavefunction
	\begin{equation}
		\psi_{\text{ho}}(r) = \left(\frac{\alpha}{\pi}\right)^{3/4} e^{-\alpha r^2/2},
	\end{equation}
	where the frequency $\omega=\sqrt{\kappa}$ determines the width parameter
	$\alpha=\omega=(6\sqrt{\pi}\sigma^3)^{-1/2}$. The corresponding ground state energy is
	\begin{equation}\label{eq:hoground}
		E_{\text{ho}} = \frac{3}{2}\omega + V_0 = \frac{3}{2(6\sqrt{\pi}\sigma^3)^{1/2}}  - \frac{1}{\sqrt{\pi}\sigma}.
	\end{equation}
	
	\begin{theorem}[Exact First-Order Correction]
		The first-order perturbative correction using the harmonic oscillator reference state is
		\begin{equation}
			\Delta E = -\frac{2}{(6\sqrt{\pi})^{1/4} \pi^{1/2} \sigma^{3/4}} \cdot \frac{1}{\sqrt{1 + 4(6\sqrt{\pi})^{-1/2} \sigma^{1/2}}}
			+ \frac{1}{\sqrt{\pi}\sigma} - \frac{(6\sqrt{\pi})^{1/2}}{8\sqrt{\pi} \sigma^{3/2}} .
		\end{equation}
	\end{theorem}
	
	\begin{proof}
		We use the notation $\langle O \rangle_{\text{ho}} := \langle \psi_{\text{ho}} | O | \psi_{\text{ho}} \rangle$.
		The first-order correction is
		\[
		\Delta E = \langle V - V_{\text{ho}} \rangle_{\text{ho}}
		= -\left\langle \frac{\erf{r/(2\sigma)}}{r} \right\rangle_{\text{ho}} + \frac{1}{\sqrt{\pi}\sigma}
		- \frac{1}{12\sqrt{\pi}\sigma^3} \langle r^2 \rangle_{\text{ho}},
		\]
		where
		\[
		\langle r^2 \rangle_{\text{ho}} = \frac{3}{2\alpha} = \frac{3}{2}(6\sqrt{\pi}\sigma^3)^{1/2}.
		\]
		The expectation value of the regularized Coulomb term is
		\[
		\left\langle \frac{\erf{r/(2\sigma)}}{r} \right\rangle_{\text{ho}}
		= 4\pi \left(\frac{\alpha}{\pi}\right)^{3/2} \int_0^\infty \erf{\frac{r}{2\sigma}}\, r\, e^{-\alpha r^2}\, dr .
		\]
		With $t=r/(2\sigma)$ and the closed integral
		\[
		\int_0^\infty t \, \erf t \, e^{-p t^2}\, dt = \frac{1}{2p\sqrt{1+p}}, \quad p = 4\alpha\sigma^2,
		\]
		we obtain
		\[
		\left\langle \frac{\erf{r/(2\sigma)}}{r} \right\rangle_{\text{ho}}
		= 2\sqrt{\frac{\alpha}{\pi}}\,\frac{1}{\sqrt{1 + 4\alpha\sigma^2}},
		\qquad
		\langle V\rangle_{\text{ho}} = -\,2\sqrt{\frac{\alpha}{\pi}}\,\frac{1}{\sqrt{1 + 4\alpha\sigma^2}} .
		\]
		Substituting $\alpha = (6\sqrt{\pi}\sigma^3)^{-1/2}$ yields
		\[
		\langle V\rangle_{\text{ho}}
		= -\,2(6\sqrt{\pi})^{-1/4} \pi^{-1/2} \sigma^{-3/4} \cdot \frac{1}{\sqrt{1 + 4(6\sqrt{\pi})^{-1/2} \sigma^{1/2}}}.
		\]
		The quadratic term is
		\[
		\frac{1}{12\sqrt{\pi}\sigma^3} \langle r^2 \rangle_{\text{ho}}
		= \frac{1}{12\sqrt{\pi}\sigma^3} \cdot \frac{3}{2\alpha}
		= \frac{(6\sqrt{\pi})^{1/2}}{8\sqrt{\pi}}\, \sigma^{-3/2}.
		\]
		Collecting the three terms gives the stated $\Delta E$.
	\end{proof}
	
	\begin{proposition}[First-Order Homotopy Energy Estimate]
		\begin{equation}\label{eq:horef}
			E_{\mathrm{total}}= \frac{1}{\sqrt{\sigma^3}} \cdot \frac{3}{4 \sqrt{6} \sqrt[4]{\pi}}
			-\frac{2}{ (6\sqrt{\pi})^{1/4} \pi^{1/2} \sigma^{3/4}} \cdot \frac{1}{\sqrt{1 + 4(6\sqrt{\pi})^{-1/2} \sigma^{1/2}}}.
		\end{equation}
		Equivalently, with $\omega=(6\sqrt\pi)^{-1/2}\sigma^{-3/2}$,
		$E_{\rm total}=\tfrac34\omega-2\sqrt{\omega/\pi}\,(1+4\omega\sigma^2)^{-1/2}$.
	\end{proposition}
	
	\begin{proof}
		$E_{\rm total}=E_{\rm ho}+\Delta E$. The constants $\pm V_0$ cancel, and
		$\tfrac32\omega-\tfrac34\omega=\tfrac34\omega$, where $\tfrac34\omega$ is the
		$\langle T\rangle_{\rm ho}$ contribution. By Lemma~\ref{lem:homomono}, $E_{\rm total}$ is an upper
		bound on the ground-state energy; it is the Rayleigh quotient of $\psi_{\rm ho}$.
	\end{proof}
	
	\begin{lemma}[Monotone homotopy]\label{lem:homomono}
		Let $V_{\rm ho}$ be as in eq.~\eqref{eq:Vho}. Then $V(r)\le V_{\rm ho}(r)$ for all $r\ge0$. Hence
		$\mathcal E(\lambda):=\inf\sigma(H(\lambda))$ is nonincreasing and concave in $\lambda$, continuous
		on $[0,1]$, and
		\[
		\mathcal E(1)\le E_{\rm ho}+\Delta E=E_{\rm total}.
		\]
	\end{lemma}
	
	\begin{proof}
		Let $f(x)=\erf x-\tfrac{2}{\sqrt\pi}(x-x^3/3)$. Then $f(0)=0$ and
		$f'(x)=\tfrac{2}{\sqrt\pi}(e^{-x^2}-1+x^2)\ge0$, so $f\ge0$. With $x=r/(2\sigma)$ and
		division by $r$ this is $V\le V_{\rm ho}$. Therefore $\partial_\lambda H=V-V_{\rm ho}\le0$ and
		$\mathcal E$ is nonincreasing. $\mathcal E(\lambda)$ is an infimum of affine functions of
		$\lambda$, hence concave and upper semicontinuous; concavity gives the matching lower bound at the
		endpoints, so $\mathcal E$ is continuous on $[0,1]$. Finally, $\psi_{\rm ho}$ is an admissible trial
		function for $H(1)$, so
		$\mathcal E(1)\le\langle\psi_{\rm ho}|H(1)|\psi_{\rm ho}\rangle=E_{\rm ho}+\Delta E$,
		which is the tangent-line bound for a concave function.
	\end{proof}
	
	
	\begin{remark}[Spectral gap of $K$]\label{rem:gap}
		Here $k_\ast=\sqrt{-2E_0}$ denotes the exact eigenvalue parameter, not the minimizer of a trial
		energy. At $k=k_\ast$, $\mu_0=1$ is a simple eigenvalue of $\tilde K$ and $\mu_1<\mu_0$
		by Prop.~\ref{prop:iterconv}(a),(b). The same conclusion follows independently from
		the Birman--Schwinger principle: for $E=-k^2/2<0$, the number of eigenvalues of
		$\tilde K_k$ larger than $1$ equals the number of eigenvalues of $H$ below $E$, and the
		multiplicity of the eigenvalue $1$ equals the multiplicity of $E$ in the spectrum of $H$.
		Since $E_\ast=-k_\ast^2/2$ is the simple lowest eigenvalue of $H$, exactly one eigenvalue
		of $\tilde K_{k_\ast}$ equals $1$ and all others lie strictly below $1$. Hence
		$\rho=\mu_1<1$. Moreover $\mu_1=1/g_1$, where $g_1>1$ is the coupling constant for which
		the first excited level of $-\tfrac12\partial_r^2-g\,m(r)$ reaches the energy $E_\ast$;
		the rate is thus qualitative and is not an explicit function of $E_1-E_\ast$.
	\end{remark}

\section{Convergence analysis}\label{sec:convergence}

Throughout, $k_\ast=\sqrt{-2E_0}$ is the exact eigenvalue parameter of the ground state, not the
minimizer of a trial energy; $K_k=SM$, $\tilde K_k=M^{1/2}SM^{1/2}$, $A_k:=SM^{1/2}$, and
$\phi_0>0$ denotes the ground-state function normalized in $L^2$.

\begin{theorem}[Convergence of Fredholm Iteration]\label{th:conv}
	Let $k=k_\ast=\sqrt{-2E_0}$, $E_\ast=-k_\ast^2/2$, $y_0(r)=re^{-k_\ast r}$ and
	$y_{j+1}=K_{k_\ast}y_j$. Then the spectrum of $\tilde K_{k_\ast}=M^{1/2}SM^{1/2}$ is
	\begin{equation}\label{eq:spectrum}
		\mu_0(k_\ast)=1>\mu_1(k_\ast)\ge\mu_2(k_\ast)\ge\dots\ge0,
	\end{equation}
	and there is a constant $C=C(\sigma,k_\ast)$ such that the normalized iterates satisfy, in
	$X=L^2$ and in $X=H^1$,
	\begin{equation}\label{eq:geomrate}
		\left\|\frac{y_j}{\|y_j\|_{L^2}}-\phi_0\right\|_{X}\le C\,\rho^{\,j},
		\qquad \rho:=\mu_1(k_\ast)<1 .
	\end{equation}
\end{theorem}

\begin{proof}
	\emph{Step 1: spectral structure.}
	By Prop.~\ref{prop:iterconv}(a), $\tilde K_{k}$ is compact, self-adjoint and non-negative;
	its largest eigenvalue $\mu_0(k)=\|\tilde K_k\|$ is simple with eigenvector $\psi_0>0$, and
	\[
	\mu_0(k)>\mu_1(k)\ge\mu_2(k)\ge\dots\ge0 .
	\]
	By Prop.~\ref{prop:iterconv}(b), $\mu_0(k_\ast)=1$. Hence
	\begin{equation}\label{eq:rho}
		\rho:=\mu_1(k_\ast)<\mu_0(k_\ast)=1 .
	\end{equation}
	The strict inequality in \eqref{eq:rho} is precisely the simplicity of $\mu_0$ stated in
	Prop.~\ref{prop:iterconv}(a); no further spectral information is required at this point.
	
	Since $K_{k_\ast}\phi_0=\phi_0$ and
	\[
	\tilde K_{k_\ast}\bigl(M^{1/2}\phi_0\bigr)=M^{1/2}K_{k_\ast}\phi_0=M^{1/2}\phi_0
	\]
	with $M^{1/2}\phi_0>0$, uniqueness of the positive eigenvector (Jentzsch--Perron--Frobenius)
	gives
	\begin{equation}\label{eq:psi0}
		\psi_0=\frac{M^{1/2}\phi_0}{\nu},\qquad \nu:=\|M^{1/2}\phi_0\|_{L^2},
	\end{equation}
	and therefore $A\psi_0=\phi_0/\nu$, where $A:=SM^{1/2}$.
	
	\emph{Step 2: non-zero overlap.}
	Set $u_j:=M^{1/2}y_j$, so that $u_{j+1}=\tilde K_{k_\ast}u_j$ and $y_{j+1}=Au_j$. Then
	\[
	c_0:=\langle u_0,\psi_0\rangle
	=\int_0^\infty m(r)^{1/2}\,r\,e^{-k_\ast r}\,\psi_0(r)\,dr>0,
	\]
	because the integrand is strictly positive; that is, $y_0$ is not orthogonal to $\phi_0$ in the
	$m$-weighted inner product.
	
	\emph{Step 3: geometric convergence.}
	Write $u_0=c_0\psi_0+r_0$ with $r_0\perp\psi_0$ and
	$\|r_0\|_{L^2}^2=\|u_0\|_{L^2}^2-c_0^2$. Since $\tilde K\psi_0=\psi_0$ and
	$\|\tilde K|_{\psi_0^\perp}\|=\mu_1=\rho$,
	\[
	u_j=c_0\psi_0+\tilde K^{j}r_0,\qquad
	\bigl\|\tilde K^{j}r_0\bigr\|_{L^2}\le\rho^{\,j}\|r_0\|_{L^2}.
	\]
	Applying $A$ and using \eqref{eq:psi0},
	\begin{equation}\label{eq:yjdecomp}
		y_{j+1}=a\,\phi_0+e_j,\qquad a:=\frac{c_0}{\nu}>0,\quad
		\|e_j\|_{X}\le\|A\|_{L^2\to X}\,\rho^{\,j}\|r_0\|_{L^2}.
	\end{equation}
	
	\emph{Step 4: bounds on $A$.}
	Since $S=2\mathcal L^{-1}$ with $\mathcal L=-d^2/dr^2+k_\ast^2$ and Dirichlet condition at
	$r=0$, the resolvent estimate gives
	\[
	\|\mathcal L^{-1}\|_{L^2\to L^2}=\frac{1}{k_\ast^2},\qquad
	\|\mathcal L^{-1}\|_{L^2\to H^1}\le\frac{\sqrt2}{\min(k_\ast,k_\ast^2)},
	\]
	where the second bound follows from $\|u\|_{H^1}^2\le(k_\ast^{-4}+k_\ast^{-2})\|f\|_{L^2}^2$
	for $u=\mathcal L^{-1}f$. Since $m$ is bounded with $\|m\|_\infty=1/(\sigma\sqrt\pi)$
	(Theorem~\ref{th:bond}), $M^{1/2}$ is bounded on $L^2$ and
	\begin{equation}\label{eq:Abounds}
		\|A\|_{L^2\to L^2}\le\frac{2\|m\|_\infty^{1/2}}{k_\ast^2}=:A_0,\qquad
		\|A\|_{L^2\to H^1}\le\frac{2\sqrt2\,\|m\|_\infty^{1/2}}{\min(k_\ast,k_\ast^2)}=:A_1 .
	\end{equation}
	In particular $A$ maps $L^2$ boundedly into $H^1_0$, so \eqref{eq:yjdecomp} holds for
	$X=L^2$ and $X=H^1$ simultaneously.
	
	\emph{Step 5: normalization.}
	Let $w=a\phi_0+e$ with $\|e\|_{L^2}\le a/2$ and $\|\phi_0\|_{L^2}=1$. Then
	$\|w\|_{L^2}\ge a-\|e\|_{L^2}\ge a/2$ and $|a-\|w\|_{L^2}|\le\|e\|_{L^2}$, so
	\begin{equation}\label{eq:normest}
		\left\|\frac{w}{\|w\|_{L^2}}-\phi_0\right\|_X
		\le\frac{2\bigl(\|e\|_X+\|e\|_{L^2}\|\phi_0\|_X\bigr)}{a}.
	\end{equation}
	Choose $j_0$ such that $A_0\|r_0\|_{L^2}\rho^{\,j_0}\le a/2$; this is possible because
	$\rho<1$ by \eqref{eq:rho}. For $j\ge j_0$, \eqref{eq:yjdecomp} and \eqref{eq:normest} give
	\[
	\left\|\frac{y_{j+1}}{\|y_{j+1}\|_{L^2}}-\phi_0\right\|_X
	\le C_X\,\rho^{\,j},
	\qquad
	C_X:=\frac{2A_X\|r_0\|_{L^2}}{a}\bigl(1+\|\phi_0\|_X\bigr),
	\]
	with $A_{L^2}=A_0$ and $A_{H^1}=A_1$ from \eqref{eq:Abounds}. Replacing $C_X$ by
	$\max_{0\le j\le j_0}\bigl(\rho^{-j}\|\hat y_{j+1}-\phi_0\|_X\bigr)\vee C_X$, which is finite
	because $\hat y_{j+1}\to\phi_0$ in the span of the retained eigendirections and $\rho>0$,
	covers the finitely many $j<j_0$. This proves \eqref{eq:geomrate} in both norms.
\end{proof}

\begin{remark}[On the rate $\rho$]\label{rem:rate}
	\begin{enumerate}
		\item[(i)] \emph{Structural meaning.} $\rho=\mu_1(k_\ast)$ is the gap between the two largest
		eigenvalues of $\tilde K_{k_\ast}$, i.e.\ the spectral gap of the Fredholm operator at the exact
		wave number. The strict inequality $\rho<1$ in \eqref{eq:rho} follows from the simplicity of
		$\mu_0$ (Prop.~\ref{prop:iterconv}(a)) together with $\mu_0(k_\ast)=1$
		(Prop.~\ref{prop:iterconv}(b)); no Birman--Schwinger argument is used in the proof of
		Theorem~\ref{th:conv}. Remark~\ref{rem:gap} records the Birman--Schwinger interpretation
		$\mu_1=1/g_1$ independently, as a consistency check, and it is not needed for the convergence
		statement.
		
		\item[(ii)] \emph{No explicit formula.} $\rho$ is \emph{not} an explicit function of the
		energy gap $E_1-E_\ast$ of $H$. By Remark~\ref{rem:gap}, $\mu_1=1/g_1$ is the reciprocal
		coupling at which the first excited level of $-\tfrac12\partial_r^2-g\,m(r)$ reaches the
		energy $E_\ast$; this is a different quantity from the excitation energy. Consequently
		Theorem~\ref{th:conv} and Corollary~\ref{cor:energy} are qualitative in the rate: they
		guarantee geometric convergence with \emph{some} $\rho<1$ but do not predict its size.
		
		\item[(iii)] \emph{Dependence on $\sigma$ and $k_\ast$.} $\rho$ depends on both parameters and
		is not monotone in either a priori. Two limiting regimes are worth recording. As
		$\sigma\to0$ the kernel reduces, up to the normalisation convention of
		\eqref{eq:greenfree}, to the Birman--Schwinger kernel of the hydrogenic problem, whose
		spectrum at $k=k_\ast$ is $\{1/n\}_{n\ge1}$; hence $\rho$ tends to a value strictly below $1$
		in the Coulomb limit, consistent with the fast observed rates at small $\sigma$
		(Table~\ref{tab:energies}). In the opposite regime $k_\ast\to0$ (large $\sigma$), the bound
		state becomes weakly localised and the Fredholm problem is ill-conditioned; one expects
		$\rho\to1$ there, so the geometric bound degrades towards the continuum threshold. The
		second statement is an expectation, not a theorem, and it is consistent with the
		deterioration of the fixed-$k$ second iterate at large $\sigma$ recorded in
		Remark~\ref{rem:kopt}.
		
	\end{enumerate}
\end{remark}

The rate is computed from the Birman--Schwinger relation $\rho=1/g_1$ by a tridiagonal
generalized eigenproblem on a uniform grid with Richardson extrapolation (script \texttt{rho\_estimate.py});
successive grids change $\rho$ by less than $10^{-7}$. The value $\rho\approx0.495$ at $\sigma=0.1$
is close to the hydrogenic limit $1/2$, and $\rho$ increases monotonically with $\sigma$, as expected
from the near-degeneracy of the oscillator-like levels. The observed error ratios are smaller than
$\rho^2$, since $k$ is re-minimised at every order; the bound of Corollary~\ref{cor:energy} is therefore
consistent with the data but not sharp.

\begin{table}[h]
	\centering
	\caption{Convergence rate $\rho=\mu_1(k_\ast)$ of the Fredholm iteration at the exact wave number,
		from the Birman--Schwinger pencil $(-\tfrac12\partial_r^2+\tfrac12k_\ast^2)y=g\,m\,y$,
		$\rho=1/g_1$. Observed ratios are those of successive minimized energy errors in
		Table~\ref{tab:energies}.}
	\label{tab:rho}
	\begin{tabular}{rcccccc}
		\toprule
		$\sigma$ & $k_\ast$ & $\rho$ & $\mu_2$ & $\rho^2$ & $\Delta E_1/\Delta E_0$ & $\Delta E_2^{\rm opt}/\Delta E_1$ \\
		\midrule
		0.1 & 0.9744 & 0.4947 & 0.3283 & 0.245 & 0.037 & 0.082 \\
		0.5 & 0.8138 & 0.5093 & 0.3465 & 0.259 & 0.073 & 0.101 \\
		1   & 0.6886 & 0.5435 & 0.3785 & 0.295 & 0.10 & 0.14 \\
		5   & 0.3931 & 0.6832 & 0.5154 & 0.467 & 0.186 & 0.283 \\
		10  & 0.2942 & 0.7480 & 0.5924 & 0.560 & 0.22 & 0.34 \\
		\bottomrule
	\end{tabular}
\end{table}

\begin{corollary}[Energy of the re-minimized iterates]\label{cor:energy}
	Let $E_m(k)$ be the Rayleigh quotient of $K_k^my_0$ and $E_m^{\rm opt}=\min_kE_m(k)$. Then
	\[
	E_0\le E_m^{\rm opt}\le E_m(k_\ast)\le E_0+C_E\,\rho^{\,2m}.
	\]
\end{corollary}

\begin{proof}
	The first two inequalities are the variational principle and the definition of the minimum.
	Let $q_{E_0}(f)=\tfrac12\|f'\|^2-\int mf^2-E_0\|f\|^2$ be the form of $H-E_0$. For a unit vector
	$y=\phi_0+\delta$, $q_{E_0}(\phi_0,\cdot)=0$ gives
	$E(y)-E_0=q_{E_0}(\delta)\le\max(\tfrac12,|E_0|)\|\delta\|_{H^1}^2$. Apply Theorem~\ref{th:conv}
	with $X=H^1$.
\end{proof}

\begin{remark}\label{rem:kopt}
	For $k\neq k_\ast$ the iteration still converges geometrically, with rate
	$\rho(k)=\mu_1(k)/\mu_0(k)$, but the limit is the dominant eigenfunction of $K_k$, which is not an
	eigenfunction of $H$ unless $\mu_0(k)=1$; its Rayleigh quotient is therefore strictly above $E_0$.
	The map $k\mapsto\tilde K_k$ is continuous in operator norm (the kernel depends continuously on
	$k$ in the Hilbert--Schmidt norm), so $\mu_0(k)$ and the gap $\mu_0(k)-\mu_1(k)$ are continuous in
	$k$; the gap stays positive on a neighbourhood of $k_\ast$ and $\rho(k)<1$ there. Consequently the
	fixed-$k$ iterates alone need not improve the energy: at $\sigma=1$ with $k=0.5545$,
	$E_1=-0.23618$ but $E_2=-0.23579$ (exact $-0.23708$). Re-minimizing over $k$ at each order, as in
	Corollary~\ref{cor:energy} and Sec.~\ref{sec:numer}, gives $E_2^{\rm opt}=-0.23695$ (at $k=0.611$).
\end{remark}

\section{Numerical Results}\label{sec:results}

All energies are in hartree and all lengths in bohr. Fredholm quantities are Rayleigh quotients of
the iterates $y_m$ evaluated on grids that start at $r=0$ (Sec.~\ref{sec:numer}); the exact
reference is the shooting/Numerov energy (Sec.~\ref{sec:shooting}). The ratios $E/E_{\min}$ use
$E_{\min}=\min V_\sigma=-1/(\sigma\sqrt\pi)$ (Corollary~\ref{corr:lower_bound}).

\subsection{Homotopic energy estimate}

The harmonic-oscillator reference energies are $E_{\rm ho}=-0.10422$ for $\sigma=1$ and
$E_{\rm ho}=+8.9036$ for $\sigma=0.1$. The positive root of eq.~\eqref{eq:hoground},
\begin{equation}
	\sigma_0=\frac{3\sqrt\pi}{8}\approx0.66467,
\end{equation}
is the screening length below which $E_{\rm ho}>0$, so the bare oscillator is unusable in the
strong-coupling regime. The first-order homotopy energies of eq.~\eqref{eq:horef} are
$E_{\rm total}=-0.18876$ for $\sigma=1$ and $E_{\rm total}=+4.2901$ for $\sigma=0.1$. The
positive root of eq.~\eqref{eq:horef} solves the cubic
$c\,t^3+4c^2t^2=64/(9\pi)$ in $t=\sigma^{-1/2}$, $c=(6\sqrt\pi)^{-1/2}$, and equals
$\sigma_h\approx0.3846$. By Lemma~\ref{lem:homomono}, $E_{\rm total}$ is an upper bound on the
ground-state energy for every $\sigma$; below $\sigma_h$ it is positive and therefore uninformative.

\begin{table}[h]
	\centering
	\caption{Harmonic-oscillator reference $E_{\rm ho}$ and first-order homotopy energy $E_{\rm total}$
		against the exact ground-state energy (relative errors in parentheses).}\label{tab:homo}
	\begin{tabular}{lccc}
		\hline
		$\sigma$ & exact & $E_{\rm ho}$ & $E_{\rm total}$ \\
		\hline
		0.1  & $-0.47478$ & $+8.9036$        & $+4.2901$ \\
		0.5  & $-0.33114$ & $+0.1726$        & $-0.1185$ (64\%) \\
		1    & $-0.23708$ & $-0.1042$ (56\%) & $-0.1888$ (20\%) \\
		5    & $-0.07726$ & $-0.0717$ (7.2\%) & $-0.0760$ (1.6\%) \\
		10   & $-0.04327$ & $-0.0419$ (3.2\%) & $-0.0430$ (0.55\%) \\
		\hline
	\end{tabular}
\end{table}

The first-order correction reduces the oscillator error by a factor $\approx2.7$ at $\sigma=1$ and
$\approx4.5$--$5.8$ at $\sigma=5,10$. The homotopy estimate is thus quantitatively useful only for
$\sigma\gtrsim5$, where the quadratic expansion of $V_\sigma$ is accurate over the extent of the
state; there it is as good as the first Fredholm iterate ($0.55\%$ against $0.60\%$ at $\sigma=10$).
The weakness at small $\sigma$ is mostly the fixed width $\alpha=\omega$ inherited from the Taylor
expansion: the same Gaussian with optimized width gives errors of $12.5\%$, $4.0\%$, $1.7\%$,
$0.16\%$ and $0.05\%$ at $\sigma=0.1,0.5,1,5,10$.

\subsection{Variational energy estimate}

\begin{figure}[h]
	\centering
	\begin{tabular}{ll}
		A & B \\
		\includegraphics[width=0.5\linewidth]{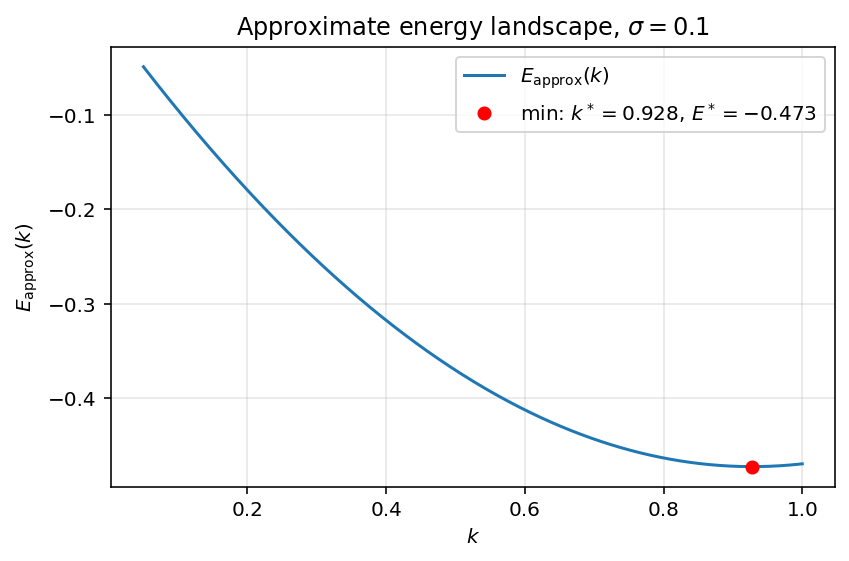} &
		\includegraphics[width=0.5\linewidth]{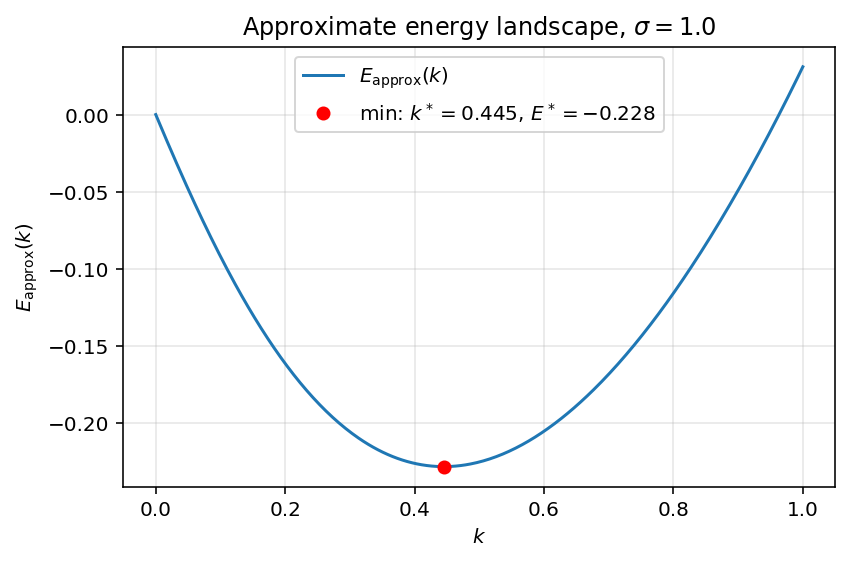}
	\end{tabular}
	\caption{Trial approximate energy landscape}
	A -- $\sigma=0.1$; B -- $\sigma=1.0$
	\label{fig:energyy0}
\end{figure}

\begin{figure}[h]
	\centering
	\begin{tabular}{ll}
		A & B \\
		\includegraphics[width=0.5\linewidth]{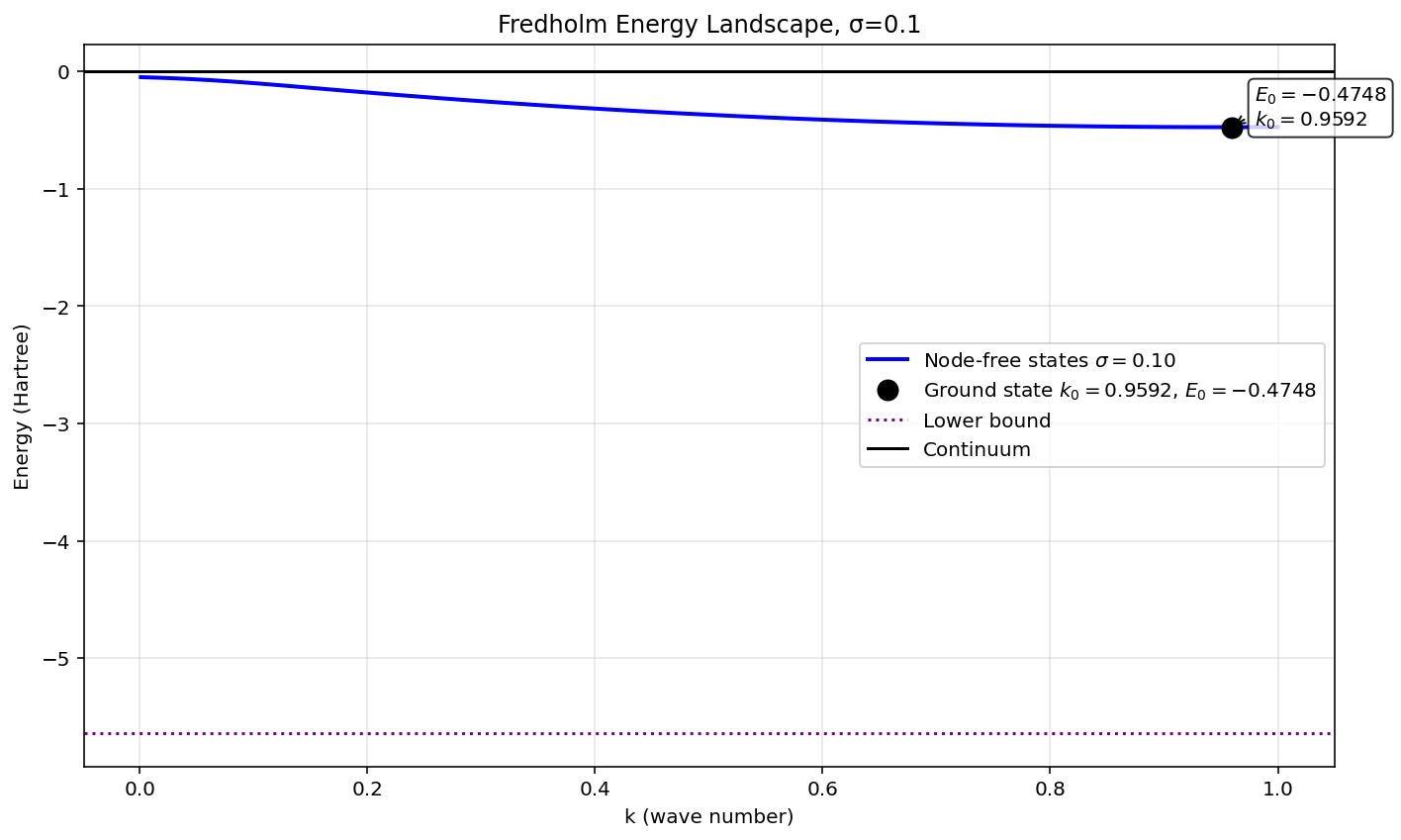} &
		\includegraphics[width=0.5\linewidth]{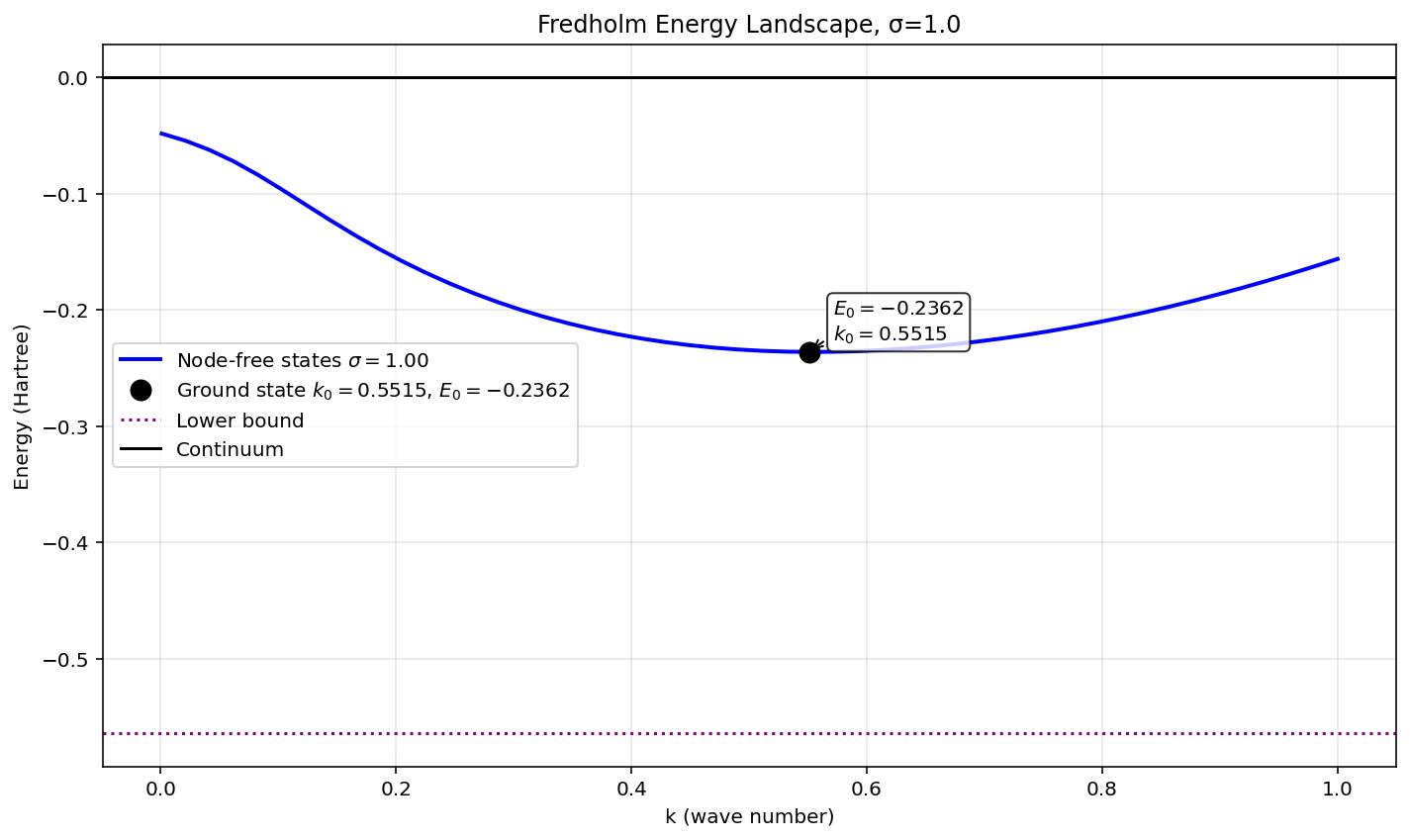}
	\end{tabular}
	\caption{Fredholm approximate energy landscape}
	A -- $\sigma=0.1$; B -- $\sigma=1.0$
	\label{fig:energyy1}
\end{figure}

The zeroth iterate is obtained by minimizing the analytic energy functional~\eqref{eq:E0}, which is
strictly convex in $k$ (Lemma~\ref{lem:lyapunov}). For weak screening ($\sigma=0.1$) it gives
$k_0=0.928$ and $E_0=-0.4729$, i.e.\ $0.39\%$ above the exact energy; the first Fredholm iterate,
minimized over $k$, gives $k_1=0.955$ and $E_1=-0.4747$ ($0.014\%$ above). For moderate screening
($\sigma=1$), $k_0=0.445$ and $E_0=-0.2284$ ($3.7\%$ above the exact $-0.2371$), while
$E_1=-0.2362$ at $k_1=0.554$ ($0.38\%$ above); the optimal wave number shifts by $25\%$. The first
iterate thus reduces the zeroth-order error by factors of $28$, $14$, $9.7$, $5.4$ and $4.6$ at
$\sigma=0.1,0.5,1,5,10$. The zeroth-order functional is fast and qualitatively correct, which makes
it a good initial guess; it is not quantitatively adequate once $\sigma\gtrsim0.5$.

\begin{table}[h]
	\centering
	\caption{Ground-state energy for the Fredholm iterates, each minimized over its own $k$ except
		the column $E_2(k_1)$, which uses the first-iterate minimizer. Relative errors (\%) against the
		exact value are given below each energy.}\label{tab:energies}
	\begin{tabular}{lcccccc}
		\hline
		$\sigma$ & reference & $E_0$ & $E_1$ & $E_2(k_1)$ & $E_2$ ($k_2$ re-min.) \\
		\hline
		0.1 & $-0.474775$ & $-0.472933$ & $-0.474709$ & $-0.474723$ & $-0.474770$ \\
		&             & 0.39        & 0.014       & 0.011       & 0.0009 \\
		0.5 & $-0.331141$ & $-0.321729$ & $-0.330458$ & $-0.330258$ & $-0.331072$ \\
		&             & 2.8         & 0.21        & 0.27        & 0.021 \\
		1   & $-0.237075$ & $-0.228396$ & $-0.236181$ & $-0.235789$ & $-0.236947$ \\
		&             & 3.7         & 0.38        & 0.54        & 0.054 \\
		5   & $-0.077256$ & $-0.074670$ & $-0.076775$ & $-0.076429$ & $-0.077120$ \\
		&             & 3.3         & 0.62        & 1.07        & 0.18 \\
		10  & $-0.043271$ & $-0.042074$ & $-0.043011$ & $-0.042796$ & $-0.043184$ \\
		&             & 2.8         & 0.60        & 1.10        & 0.20 \\
		\hline
	\end{tabular}
	
	For reference is taken as the shooting/Numerov baseline.
\end{table}

All energies lie above the exact value, as required by the variational principle. 
The optimal wave numbers are $k_0=0.928,0.612,0.445,0.174,0.110$, $k_1=0.955,0.717,0.555,0.237,0.153$ and $k_2=0.965,0.763,0.611,0.279,0.184$, 
against the reference $k=0.974,0.814,0.689,0.393,0.294$ for
$\sigma=0.1,0.5,1,5,10$.

\subsection{Ground state calculation}

\begin{figure}
	\centering
	\begin{tabular}{ll}
		A & B \\
		\includegraphics[width=0.5\linewidth]{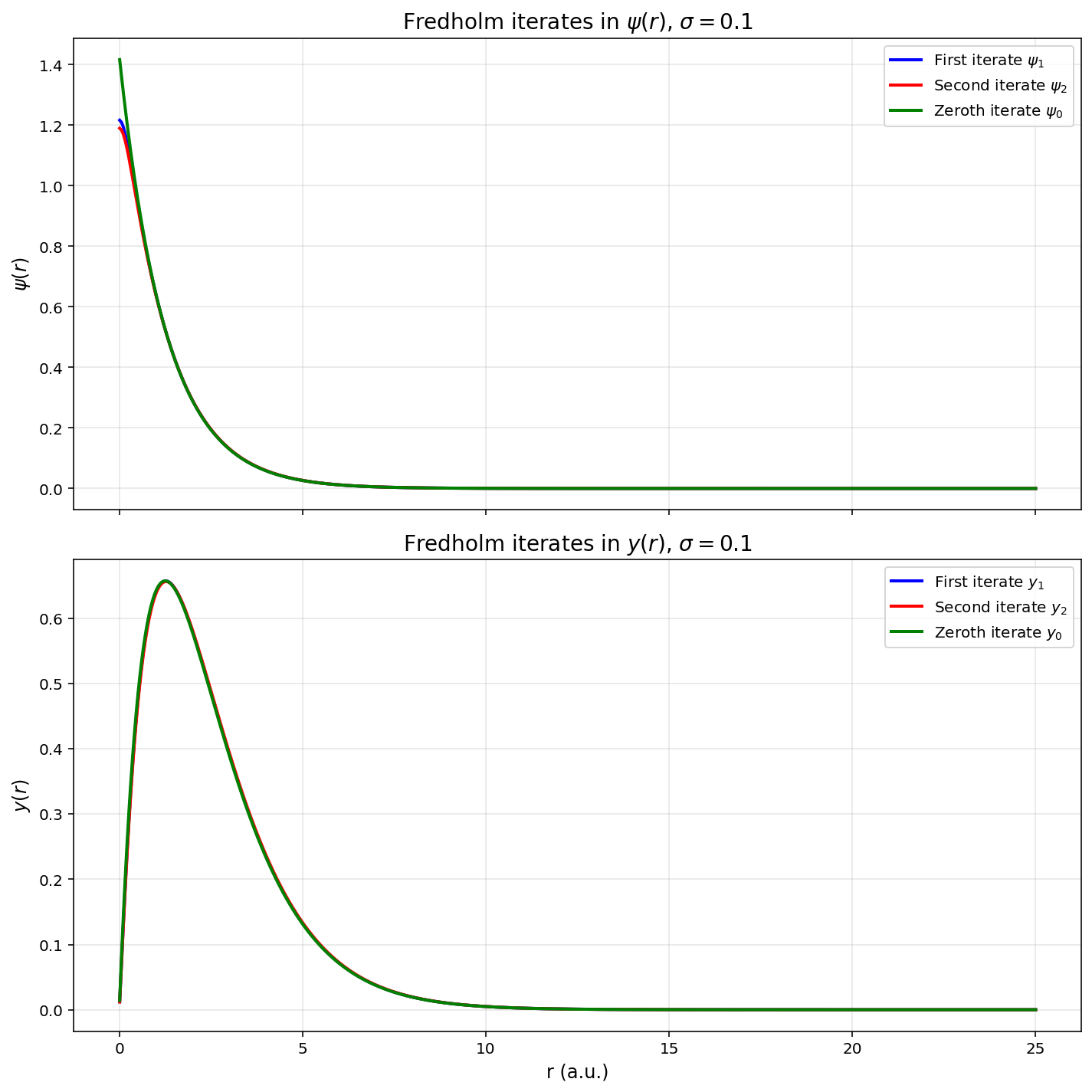} & \includegraphics[width=0.5\linewidth]{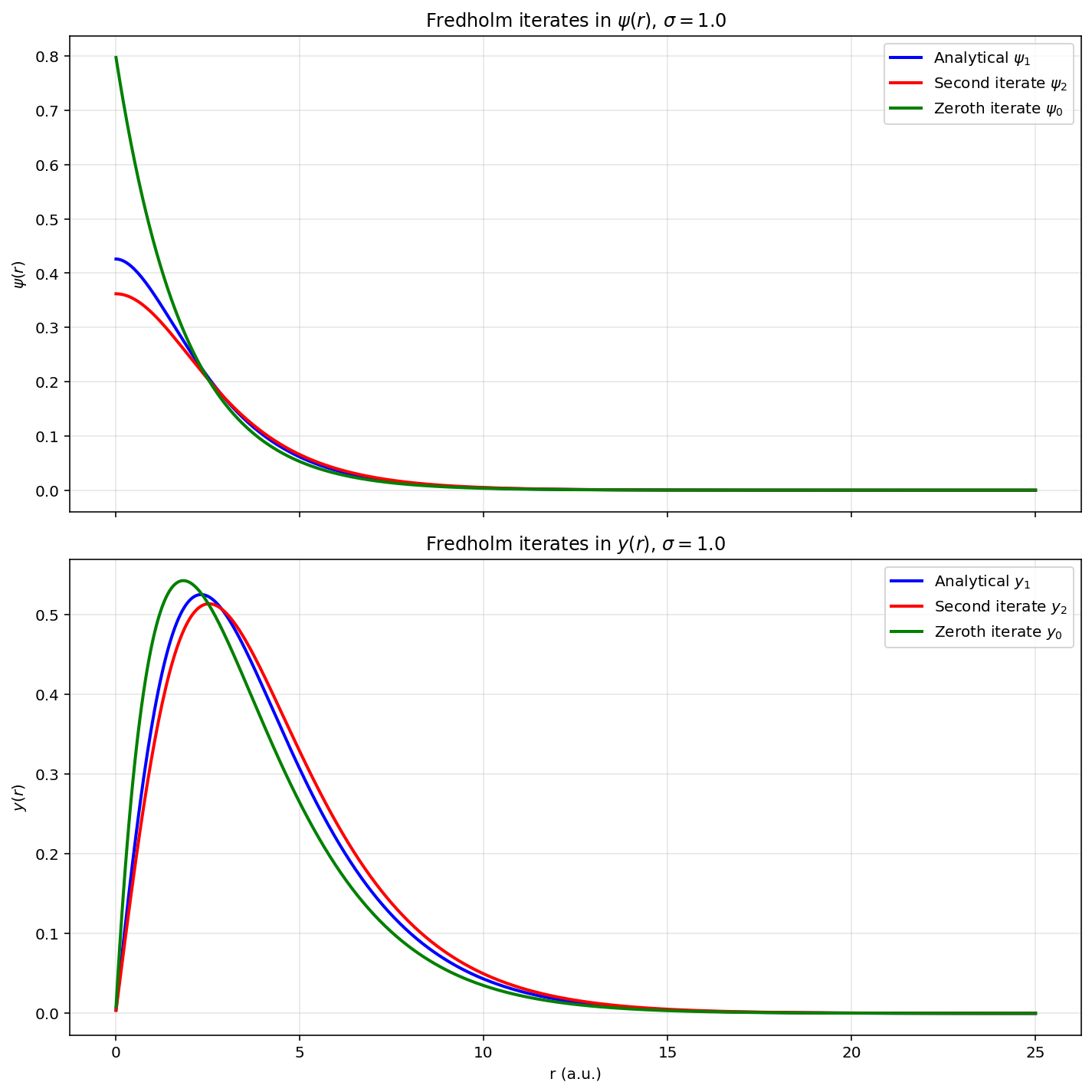}
	\end{tabular}
	\caption{Fredholm iteration steps for two screening parameters}
	A-- $\sigma=0.1$; B -- $\sigma=1.0$
	\label{fig:fedholmiter1}
\end{figure}

\begin{table}[h]
	\centering
	\caption{Mean radius $\langle r\rangle$ and rms radius $\sqrt{\langle r^2\rangle}$ (bohr) of the
		iterates and of the exact ground state; all iterates are node-free.}\label{tab:radii}
	\begin{tabular}{llccccc}
		\hline
		$\sigma$ & & exact & $y_0$ & $y_1$ & $y_2(k_1)$ & $y_2(k_2)$ \\
		\hline
		0.1 & $\langle r\rangle$   & 1.579  & 1.617  & 1.587  & 1.598  & 1.582 \\
		& $\sqrt{\langle r^2\rangle}$ & 1.816 & 1.867 & 1.829 & 1.839 & 1.820 \\
		0.5 & $\langle r\rangle$   & 2.234  & 2.451  & 2.274  & 2.367  & 2.245 \\
		& $\sqrt{\langle r^2\rangle}$ & 2.516 & 2.830 & 2.584 & 2.678 & 2.537 \\
		1   & $\langle r\rangle$   & 3.035  & 3.371  & 3.101  & 3.302  & 3.055 \\
		& $\sqrt{\langle r^2\rangle}$ & 3.379 & 3.892 & 3.499 & 3.703 & 3.419 \\
		5   & $\langle r\rangle$   & 7.928  & 8.614  & 8.081  & 9.083  & 7.981 \\
		& $\sqrt{\langle r^2\rangle}$ & 8.685 & 9.946 & 9.052 & 10.076 & 8.844 \\
		10  & $\langle r\rangle$   & 12.695 & 13.585 & 12.888 & 14.782 & 12.758 \\
		& $\sqrt{\langle r^2\rangle}$ & 13.862 & 15.686 & 14.430 & 16.375 & 14.127 \\
		\hline
	\end{tabular}
\end{table}

For moderate screening ($\sigma=1$) the first iterate gives $E_1=-0.2362$ at $k_1=0.5545$,
which is $41.9\%$ of $E_{\min}=-0.5642$ (the exact state sits at $42.0\%$). Its mean radius is
$3.10$ against the exact $3.04$ ($+2.2\%$). The second iterate at the same $k$ is
less accurate in energy ($-0.2358$) and in size ($\langle r\rangle=3.30$, $+8.8\%$): at fixed $k$
the iteration approaches the dominant eigenfunction of $K_k$, which is not the ground state
unless $\mu_0(k)=1$ (Remark~\ref{rem:kopt}). Re-minimizing $k$ ($k_2=0.611$) restores
$E_2=-0.2369$ and $\langle r\rangle=3.055$ ($+0.7\%$). The same pattern appears at larger $\sigma$:
at $\sigma=5$ the fixed-$k$ second iterate overestimates $\langle r\rangle$ by $14.6\%$ and at
$\sigma=10$ by $16.4\%$, while the re-minimized one stays within $0.7\%$.

In the strong-coupling case ($\sigma=0.1$) all three iterates have nearly identical radii
(within $2.4\%$ of the exact $1.58$ for $y_0$ and within $0.5\%$ for $y_1$), and the corresponding
energies agree with the exact value to $0.39\%$, $0.014\%$ and $0.0009\%$ for $E_0,E_1,E_2$. A single
Fredholm correction therefore brings the error below $10^{-3}$ hartree at every $\sigma$ tested
($6.5\times10^{-5}$ at $\sigma=0.1$ to $8.9\times10^{-4}$ at $\sigma=1$), and each re-minimized
iterate improves on the previous one. The kinetic energy computed from the analytic derivative of
$y_1$ agrees with a finite-difference derivative to $\le10^{-6}$ hartree on a grid of $2\times10^4$
points.

\begin{figure}[h]
	\centering
	\includegraphics[width=1.0\linewidth]{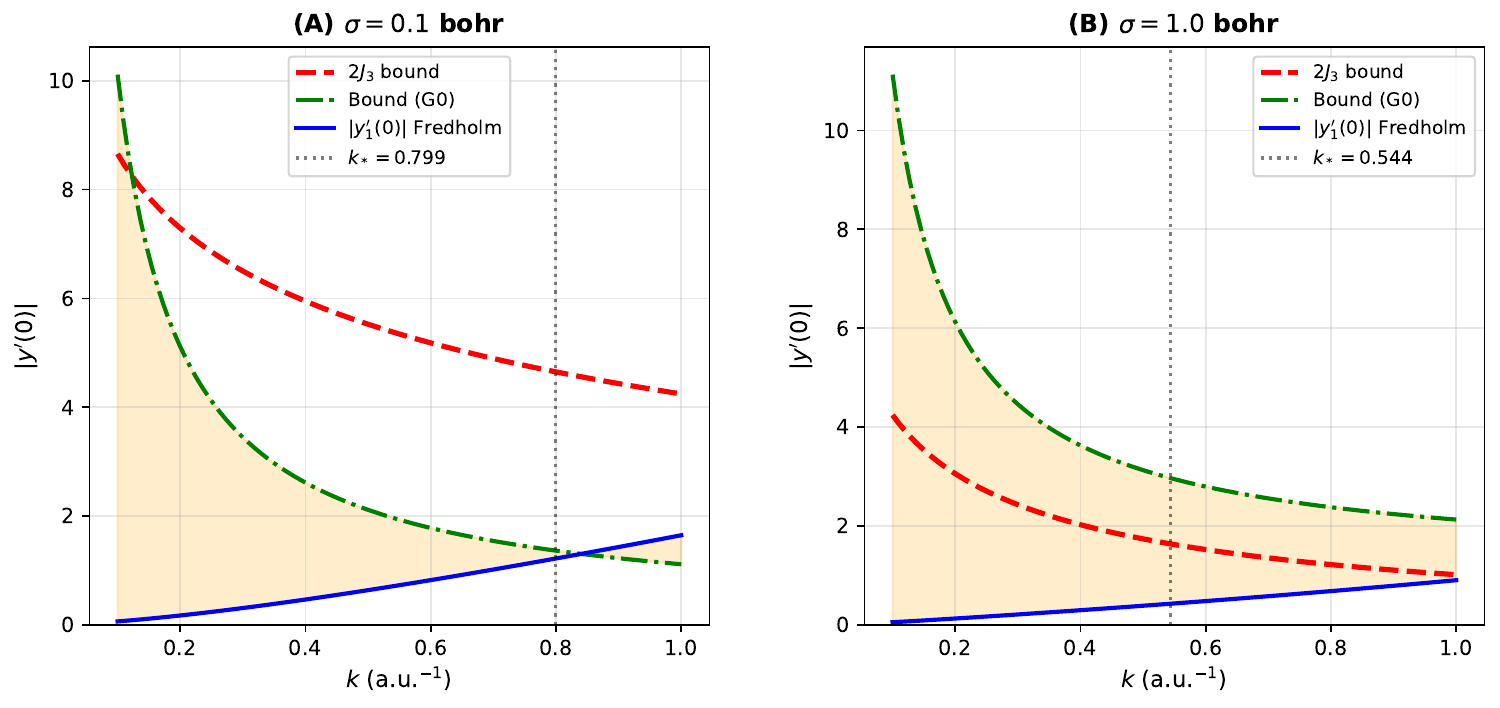}
	\caption{Gradient at the origin as a function of the trial wavenumber $k$; A -- $\sigma=0.1$;
		B -- $\sigma=1.0$. Cyan: unnormalized $y_1'(0)=e^{x^2}\mathrm{erfc}(x)/k$, $x=2k\sigma$, bounded by
		$G_0=2\sigma/\sqrt\pi+1/k$ (green dash-dotted). Blue: normalized $\hat y_1'(0)$, bounded by
		$2\|\hat y_0\|_\infty J_3/\|K\hat y_0\|$ (red dashed, rigorous). Gray dotted: $2J_3$, which held
		numerically on $k\in[0.1,1]$ but is not proved for the iterate. The vertical dotted line marks the
		exact wave number $k_\ast$.}
	\label{fig:y1primebounds}
\end{figure}

The bounds are compared with the gradients in Fig.~\ref{fig:y1primebounds}. At the first-iterate
optimum $k_1$ the unnormalized gradient $y_1'(0)=0.855$ ($\sigma=0.1$) and $0.721$ ($\sigma=1$)
lies below $G_0=1.160$ and $2.932$ (margins $1.36$ and $4.07$); the normalized gradient $1.547$ and
$0.436$ lies below the rigorous bound $3.017$ and $0.648$ (margins $1.95$ and $1.49$). For the exact
ground state the bound $y'(0)\le2\|y\|_\infty J_3$ exceeds $y'(0)$ by factors $2.04,\,1.82,\,1.87,\,2.22,\,2.47$
and the Cauchy--Schwarz bound by $3.89,\,2.39,\,2.34,\,2.97,\,3.56$ for $\sigma=0.1,0.5,1,5,10$.

\subsection{Comparison with harmonic-oscillator perturbation theory}

For moderate screening ($\sigma=1$) the oscillator energy $E_{\rm ho}=-0.104$ underestimates the
binding energy by $56\%$ and first-order homotopy reduces the error to $20\%$
($E_{\rm total}=-0.189$) against the exact $-0.237$ (Table~\ref{tab:homo}); the first Fredholm
iterate is within $0.38\%$. For weak screening ($\sigma=0.1$) the oscillator is unbound
($E_{\rm ho}=+8.90$, a relative error of $1975\%$) and first order ($+4.29$) is still
unbound ($1004\%$ error), while the Fredholm iterates correctly give a bound state at
$-0.4747$. Here the potential is deep and nearly Coulombic ($V_0=-5.64$, with the ground state at
$8.4\%$ of this depth), so the parabolic approximation is not applicable. For $\sigma\ge5$,
however, first-order homotopy is competitive with the first Fredholm iterate. The advantage of the
Fredholm approach is therefore concentrated at small and moderate $\sigma$, where it remains accurate
and variationally bounded across the whole range tested.

\subsection{Comparison with direct integration}
\label{sec:shooting}

The algorithms are described in Sec.~\ref{sec:shooting-impl}. Table~\ref{tab:shooting} compares the
reference energies with the Fredholm results.

\begin{table}[ht]
	\centering
	\caption{Ground-state energy (hartree): reference from shooting and Numerov matching, and the
		Fredholm iterates $E_1$ and $E_2$ (each minimized over $k$).}
	\label{tab:shooting}
	\begin{tabular}{lccc}
		\hline
		$\sigma$ & reference (shooting = Numerov) & $E_1$ & $E_2$ \\
		\hline
		0.1  & $-0.47477539$ & $-0.474710$ & $-0.474771$ \\
		0.5  & $-0.33114072$ & $-0.330458$ & $-0.331072$ \\
		1    & $-0.23707528$ & $-0.236181$ & $-0.236947$ \\
		5    & $-0.07725572$ & $-0.076775$ & $-0.077120$ \\
		10   & $-0.04327087$ & $-0.043011$ & $-0.043184$ \\
		\hline
	\end{tabular}
\end{table}

Shooting and Numerov agree with each other to $\lesssim10^{-12}$ and are insensitive to $r_{\max}$
and to halving $h$; only a loose tolerance (RK45 at $\mathrm{rtol}=10^{-3}$) degraded the energy, by
about $3\times10^{-5}$. A second-order finite-difference diagonalization deviates from them by
$2.7\times10^{-7}$ at $\sigma=0.1$ (its own discretization error).

For the ground state of this one-dimensional problem, direct integration is accurate and
inexpensive: a converged energy needs $34$--$39$ ODE solves. The Fredholm approach does not
outperform it in accuracy at a fixed number of terms. Its advantages are different:
(i) $y_1$ is explicit in terms of $\mathrm{erf}$ and $\mathrm{erfc}$, so $E_1(k)$ is a smooth
one-parameter function; (ii) every minimized $E_m$ is a rigorous upper bound, and
$E_m^{\rm opt}-E_0=O(\rho^{2m})$ (Corollary~\ref{cor:energy}); (iii) the iteration comes with
a-priori bounds (Prop.~\ref{prop:E0bounds}) and a convergence theorem (Thm.~\ref{th:conv}), and the
gradient $y'(0)$ is controlled analytically. Shooting provides none of these, but it needs no
a-priori choice of $k$.

\paragraph{Caveats.}
Integrating at a fixed $k$ that is not an eigenvalue does not test a shooting method: for
$\sigma=0.1$ the exact $k=0.9744$, and at $k=0.799$ the outward solution has its maximum at
$r=0.92$ instead of the exact $1.07$. A test that fixes $y'(0)=1$ by the initial condition says
nothing about the bound on the normalized gradient, which applies to $y'(0)/\|y\|_2$. Only the
$s$-wave ground state was tested, for $\sigma\in[0.1,10]$; excited states and the near-threshold
regime ($E\to0^-$ at large $\sigma$) were not studied. 

\subsection{Comparison with a Gaussian basis}
\label{sec:gaussian}

As shown in the previous sections, the Gaussian family of functions re closely related to the structure of the problem. 
Therefore, as a second benchmark we solve the radial problem variationally in a basis of $N$ even-tempered
Gaussians $u_i(r)=r\,e^{-a_ir^2}$, $a_i=a_{\min}(a_{\max}/a_{\min})^{(i-1)/(N-1)}$. All matrix
elements are analytic. With $p=a_i+a_j$ and $b=1/(2\sigma)$,
\begin{equation}
	S_{ij}=\frac{\sqrt\pi}{4p^{3/2}},\qquad
	T_{ij}=\frac{3\sqrt\pi\,a_ia_j}{4p^{5/2}},\qquad
	V_{ij}=-\frac{b}{2p\sqrt{p+b^2}},
\end{equation}
where the last follows from $\int_0^\infty re^{-pr^2}\mathrm{erf}(br)\,dr=b/(2p\sqrt{p+b^2})$.
The lowest root of the generalized eigenproblem is an upper bound on the exact energy. Near-linear
dependence is removed by canonical orthogonalization (threshold $10^{-13}$). For each $N$ the
exponent range was chosen by minimizing the energy over twelve $(a_{\min},a_{\max})$ pairs, so the
selection does not use the reference value. The reference is a finite-difference solution with
Richardson extrapolation, whose own error is about $10^{-10}$~hartree; this sets the floor of the
comparison below.

\begin{table}[t]
	\centering
	\caption{Energy error $E-E_{\rm ref}$ (hartree) of the Fredholm iterates and of a Gaussian basis of
		$N$ functions. $E_1$ and $E_2^{\rm opt}$ are the minimized first and second Fredholm iterates.
		Entries marked $\lesssim$ are below the accuracy of the reference.}
	\label{tab:gaussian}
	\begin{tabular}{rccccc}
		\toprule
		$\sigma$ & Fredholm $E_1$ & Fredholm $E_2^{\rm opt}$ & Gauss $N{=}6$ & Gauss $N{=}10$ & Gauss $N{=}15$ \\
		\midrule
		0.1  & $6.7\times10^{-5}$ & $4.9\times10^{-6}$     & $2.5\times10^{-3}$ & $3.1\times10^{-5}$ & $1.3\times10^{-8}$ \\
		0.5  & $6.8\times10^{-4}$ & $6.9\times10^{-5}$   & $6.3\times10^{-5}$ & $1.7\times10^{-7}$ & $\lesssim10^{-10}$ \\
		1    & $8.9\times10^{-4}$ & $1.3\times10^{-4}$   & $1.3\times10^{-4}$ & $9.9\times10^{-9}$ & $\lesssim10^{-10}$ \\
		5    & $4.8\times10^{-4}$ & $1.4\times10^{-4}$   & $9.7\times10^{-5}$ & $3.4\times10^{-7}$ & $\lesssim10^{-10}$ \\
		10   & $2.6\times10^{-4}$ & $8.7\times10^{-5}$   & $1.9\times10^{-5}$ & $4.5\times10^{-8}$ & $\lesssim10^{-10}$ \\
		\bottomrule
	\end{tabular}
\end{table}

Table~\ref{tab:gaussian} shows that six Gaussians already match the re-minimized second Fredholm
iterate for $\sigma\ge0.5$, ten give errors of $10^{-7}$ or smaller, and fifteen reach the accuracy of
the reference for $\sigma\ge0.5$ (and $10^{-8}$ at $\sigma=0.1$).

\subsection{Physical Interpretation}

Our screening length is related to the HSE range-separation
parameter $\omega$ by $\sigma=1/(2\omega)$; for $\omega\approx0.11$~bohr$^{-1}$ this gives $\sigma\approx4.5$~bohr.

The computed ground states are weakly to moderately bound. Relative to the trivial lower bound
$E_{\min}=\min V_\sigma=-1/(\sigma\sqrt\pi)$ (Corollary~\ref{corr:lower_bound}), the exact energies
lie at $8.4\%$ ($\sigma=0.1$) and $42.0\%$ ($\sigma=1$), and the first iterate gives $8.4\%$ and
$41.9\%$; since $E_{\min}$ is the depth of the potential, the ratio measures the binding relative to
the potential minimum, not a sharp bound.

For $\sigma=1$ the exact state has $\langle r\rangle=3.04$~a.u.\ and
$\sqrt{\langle r^2\rangle}=3.38$~a.u. The first iterate (at $k_1=0.555$) gives $3.10$ and
$3.50$~a.u., and the second iterate at the same $k$ gives $3.30$ and $3.70$~a.u., i.e.\ it moves
away from the exact size; with $k$ re-minimized ($k_2=0.611$) it gives $3.06$ and $3.42$~a.u.
(within $1\%$ for the mean radius). For $\sigma=0.1$ the state is more compact (exact $1.58$ and
$1.82$~a.u., first iterate $1.59$ and $1.83$~a.u.), reflecting the stronger effective Coulomb
attraction as screening diminishes.

The energy errors tell the same story. The first iterate is $0.014\%$ ($\sigma=0.1$) and $0.38\%$
($\sigma=1$) above the exact energy; the second iterate at fixed $k$ is $0.011\%$ and $0.54\%$ above
it, while with $k$ re-minimized it is $0.0009\%$ and $0.054\%$ above. These errors are consistent
with the geometric rate of Theorem~\ref{th:conv}, which holds at $k=k_\ast$ only. All iterates are
node-free; this is a property of the construction (the kernel of $K_k$ is positive and $y_0>0$), not
an empirical confirmation of ground-state character.

\section{Discussion}

The present work makes three contributions.
First, we develop a Fredholm integral-equation formulation of the radial ground state for the
erf-screened Coulomb potential. The first iterate is available in closed form through $\mathrm{erf}$,
$\mathrm{erfc}$ and exponentials. At the exact wave number the iteration converges geometrically
with a rate $\rho<1$ (Theorem~\ref{th:conv}); $\rho<1$ follows from the Birman--Schwinger principle
(Remark~\ref{rem:gap}), but the rate is not an explicit function of the spectral gap. The
harmonic-oscillator homotopy connects the problem to a
solvable reference. Along the homotopy path the ground state stays simple, and its energy is nonincreasing and
concave in the homotopy parameter. Because $V_\sigma\le V_{\rm ho}$ (Lemma~\ref{lem:homomono}), the
first-order homotopy energy is a rigorous upper bound.  

Second, we use the Rayleigh quotients of the iterates as variational energies. The zeroth-order trial
energy is strictly convex in the decay parameter with a unique minimizer, and it is a Lyapunov
function for the gradient flow of that parameter (Lemma~\ref{lem:lyapunov}); this concerns the
convergence of the parameter, not of the iteration to the exact ground state. Each minimized iterate
energy is an upper bound on the exact energy, and the re-minimized energies approach it as
$\rho^{2m}$ with the iteration order $m$ (Corollary~\ref{cor:energy}). In every case we computed they
decreased monotonically with $m$, which we do not prove. Re-minimizing the decay parameter matters:
at fixed parameter the second iterate is worse than the first for $\sigma\ge0.5$
(Remark~\ref{rem:kopt}).

Third, we benchmark against other techniques - notably direct integration and Gaussian basis representation. 
Shooting with node counting and Numerov matching
agree to about $10^{-12}$ hartree (Sec.~\ref{sec:shooting}), so direct integration is an accurate and
inexpensive reference for this one-dimensional problem. The first-order homotopy estimate is useful
only at large screening length (errors of $1.6\%$ and $0.55\%$ at $\sigma=5$ and $10$) and is
positive for $\sigma<0.385$.
Numerical results demonstrate that a small
Gaussian basis set is  more accurate than the two Fredholm iterates. 

The advantages of the Fredholm approach are theoretical and structural rather than a simple gain in accuracy: closed-form iterates, guaranteed upper bounds with a geometric rate, a priori bounds on
the energy and the optimal parameter (Prop.~\ref{prop:E0bounds}), and analytic control of the
gradient at the origin. 
The zeroth-order functional supplies two quantities useful as initial data: a rigorous upper bound
$E_0(k_0^\ast)\ge E_{\rm exact}$ with relative error $0.4$--$3.7\%$ for $\sigma\in[0.1,10]$,
and the wave number $k_E=\sqrt{-2E_0(k_0^\ast)}$, which lies within $1.8\%$ of the exact $k$ and
bounds it from below. Together with $k<1$ they give the bracket $[k_E,1)$ for Numerov matching.
The trial exponent $k_0^\ast$ itself is a poor estimate of $k$ for $\sigma\ge0.5$.

\subsection{Outlook}

The present model is a single-particle radial problem with a model potential, and the tests cover the
$s$-wave ground state for $\sigma\in[0.1,10]$ only; excited states, non-zero angular momentum and the
near-threshold regime at large $\sigma$ were not studied. Relating $\sigma$ to a screening length in a
specific material, or using the solver inside a many-body method such as HSE, requires further work
that is beyond the scope of this paper and is the natural next step.

The present approach extends to other spherically symmetric, screened or short-range potentials for which the
kernel integrals and Green's functions are available in closed form, for example the Yukawa
(Debye--H\"uckel) potential and exponential-cosine screened potentials. For the present model the
first iterate is analytic and $E_1(k)$ is a smooth function of one parameter, so a complete
ground-state determination (12--53 gradient iterations, or 10--12 evaluations with Brent's method)
took $0.1$--$0.3$~s in an unoptimized Python implementation, and parameter scans in $\sigma$ are
inexpensive. Direct integration is equally cheap for the single-particle radial problem, so the
practical value of the method lies in its analytic structure, the rigorous upper bounds and the
a priori estimates, and as a starting point for problems where the kernel is known in closed form but
a differential solver is awkward.

	\section*{Declaration of Generative AI}
	Perplexity AI was used for deep research on the topic. 
	
	\bibliographystyle{elsarticle-num}
	\bibliography{erfpaper_updated}
	
	\appendix
	\section{Hypergeometric functions}
	
	The generalized hypergeometric functions are defined by the infinite hypergeometric (HG) series
	\begin{multline}
		\hsym{p}{q} ( a_1, \ldots, a_p; b_1, \ldots, b_q, x)  :=  \\ \sum\limits_{m=0}^{\infty} \frac{x^m}{\Gamma(m + 1)} \prod^{p}_{k=1} \frac{\Gamma( a_k+ m)  }{\Gamma( a_k)  } 
		\prod^{q}_{k=1} \frac{  \Gamma( b_k)}{  \Gamma( b_k+ m) } =
		\sum_{r=0}^{\infty} \frac{x^r}{r!} \frac{\prod_{j=0}^{p-1}\
			(a_j)_r}{ \prod_{j=0}^{q-1}(b_j)_r} 
	\end{multline}
	where $(a)_r$ and $(b)_r$ will denote rising factorials and $(a)_0=1$, which assumes the normalization
	$\hsym{p}{q}\left( \sim; \sim \Big| 0 \right) =1 $. 
	By convention, equal parameters in the numerator and denominator will cancel out.
	Unless stated otherwise it will be always assumed that the infinite series converge in some domain $x \in \fclass{R}{}$. 
	For example the erf function can be written as
	\[
	\erf{x} = \frac{2}{\sqrt \pi} x \hsym{1}{1} ( 1/2; 3/2; x)
	\]

	\section{Spectrum equivalence}\label{sec:proofL2}
	
	\begin{lemma}[Equivalence of Schr\"odinger and Fredholm Spectra]\label{th:equiv}
		
		Suppose that $H$ is defined on the domain of smooth $L^2$-functions with $y(0)=0$ and $y(r)\to 0$ as $r\to\infty$.
		For $E=-k^2/2<0$, let $K(E)$ be the integral operator defined by the Fredholm equation
		\[
		y(r) = \int_0^\infty G_k(r,r')\,V(r')\,y(r')\,dr',
		\]
		where $G_k$ is the free Green's function \eqref{eq:greenfree} at energy $E$. Then
		\[
		Hy = E y \text{ with } y\in L^2,\ y(0)=0,\ y(r)\to0
		\quad\Longleftrightarrow\quad
		K(E)y = y.
		\]
		In particular, $E$ is a bound-state eigenvalue of $H$ if and only if $1$ is an eigenvalue of $K(E)$ with eigenfunction $y$.
	\end{lemma}
	\begin{proof}
		\emph{($\Rightarrow$)} 
		Assume $Hy = E y$ with $y \in L^2(0,\infty)$, $y(0)=0$, and $y(r)\to 0$ as $r\to\infty$.
		Rewrite the eigenvalue equation as
		\[
		\Bigl(-\frac12 \frac{d^2}{dr^2} - E\Bigr) y(r) = - V(r)\,y(r).
		\]
		Denote the free radial operator by
		\[
		H_0\; y(r) := -\frac12\,y''(r), \qquad (H_0 - E) y = - V y.
		\]
		For $E<0$, the operator $H_0 - E$ is invertible on the radial $L^2$ space with
		Dirichlet boundary condition at $r=0$ and decaying solutions at infinity, and its inverse
		is the integral operator with kernel $G_k(r,r')$ \eqref{eq:greenfree}. Thus
		\[
		y = (H_0 - E)^{-1}(-V y)
		= \int_0^\infty G_k(r,r')\,V(r')\,y(r')\,dr' = K(E) y.
		\]
		Hence $K(E) y = y$.
		
		\emph{($\Leftarrow$)} 
		Conversely, suppose $y \in L^2(0,\infty)$ satisfies $K(E) y = y$, that is
		\[
		y(r) = \int_0^\infty G_k(r,r')\,V(r')\,y(r')\,dr'.
		\]
		Apply the differential operator $H_0 - E = -\tfrac12\,\frac{d^2}{dr^2} - E$ to both sides.
		Using that $G_k$ is the Green's function for $H_0 - E$, we have
		\begin{multline*}
			\Bigl(-\frac12 \frac{d^2}{dr^2} - E\Bigr) y(r)
			= \int_0^\infty
			\Bigl(-\frac12 \frac{\partial^2}{\partial r^2} - E\Bigr)
			G_k(r,r')\,V(r')\,y(r')\,dr' \\
			= \int_0^\infty \delta(r-r')\,V(r')\,y(r')\,dr'
			= V(r)\,y(r).
		\end{multline*}
		Therefore,
		\[
		-\tfrac12\,y''(r) + V(r)\,y(r) = E\,y(r),
		\]
		so $Hy = E y$. The construction of $G_k$ with Dirichlet behavior at $r=0$ and exponential
		decay at infinity ensures that $y(0)=0$ and $y(r)\to 0$ as $r\to\infty$ whenever
		$y \in L^2(0,\infty)$ solves the integral equation. Thus $y$ is a bound-state eigenfunction
		of $H$ at energy $E$.
		
		This establishes the equivalence between bound-state eigenvalues of $H$ and unit
		eigenvalues of the compact, self-adjoint Fredholm operator $K(E)$.
	\end{proof}
	
	\section{Derivation of the Green function}\label{sec:green}
	
	We derive the Green's function for the operator $\mathcal L := -\frac12\partial_r^2 + \frac{k^2}{2}$ on $(0,\infty)$ with Dirichlet condition at $r=0$ and decay at $\infty$, i.e.\ the solution of
	\[
	\Bigl(-\tfrac12\partial_r^2+\tfrac{k^2}{2}\Bigr) G_k(r,r') = \delta(r-r'), \quad G_k(0,r')=0, \quad \lim_{r\to\infty}G_k(r,r')=0.
	\]
	Equivalently, $G_k'' - k^2 G_k = -2\,\delta(r-r')$.
	For \(0 \le r < r'\): \(G_k(r,r') = A \sinh{(kr)}\), while for \(r > r'\): \(G_k(r,r') = B e^{-kr}\).
	
	Continuity at \(r = r'\) implies
	\[
	A \sinh{(k r')} = B e^{-k r'} \quad \Rightarrow \quad B = A \sinh{(k r')} e^{k r'}.
	\]
	Jump condition for $G''-k^2G=-2\delta$: $G_2'(r') - G_1'(r') = -2$, i.e.
	\[
	-k B e^{-k r'} - k A \cosh{(k r')} = -2
	\Rightarrow
	-k A [\sinh{(k r')} + \cosh{(k r')}] = -2.
	\]
	Since \(\sinh x + \cosh x = e^x\):
	\[
	-k A e^{k r'} = -2 \quad \Rightarrow \quad A = \frac{2\,e^{-k r'}}{k}.
	\]
	%
	Thus, with the convention of eq. \eqref{eq:greenfree}, the compact form is
	\[
	G_k(r, r') = -\frac{1}{k} e^{-k r_>} \sinh{(k r_<)},
	\]
	where \(r_< = \min{(r, r')}, \quad r_> = \max{(r, r')} \).

	\section{Exact First Iteration with \(y_0(r) = r e^{-kr}\)}\label{sec:iter}
	
	We compute the first iteration exactly for the Fredholm equation:
	\[
	y_{k+1}(r) = \frac{2}{k} e^{-k r} \int_0^r \frac{\erf{ \frac{r'}{2\sigma} }}{r'} \sinh(k r') y_{k}(r') \, dr'
	+ \frac{2}{k} \sinh(k r) \int_r^\infty \frac{\erf{ \frac{r'}{2\sigma} }}{r'} e^{-k r'} y_{k}(r') \, dr'.
	\]
	with \(y_0(r') = r' e^{-k r'}\), where
	\[
	I_1(r) = \int_0^r \frac{\erf{ \frac{r'}{2\sigma} }}{r'} \sinh(k r')   r' e^{-k r'} \, dr'
	= \int_0^r \erf{ \frac{r'}{2\sigma} } \sinh(k r') e^{-k r'} \, dr'.
	\]
	Using \(\sinh(k r') e^{-k r'} = \frac{1 - e^{-2k r'}}{2}\):
	\[
	I_1(r) = \frac{1}{2} \int_0^r \erf{ \frac{r'}{2\sigma} } dr' - \frac{1}{2} \int_0^r \erf{ \frac{r'}{2\sigma} } e^{-2k r'} dr'.
	\]
	and
	\[
	I_2(r) = \int_r^\infty \frac{\erf{ \frac{r'}{2\sigma} }}{r'} e^{-k r'} \cdot r' e^{-k r'} \, dr'
	= \int_r^\infty \erf{ \frac{r'}{2\sigma} } e^{-2k r'} dr'.
	\]
	
	\subsection{General Indefinite Integral}
	
	Consider the integral 
	\[
	\int \erf{ax} e^{bx} dx = \frac{1}{b} e^{bx} \erf{ax} - \frac{1}{b} e^{\frac{b^2}{4a^2}} \erf{ax - \frac{b}{2a}} + C,
	\]
	with \(a = 1\), \(b = -2\alpha\sigma\):
	\[
	\int \erf t e^{-2\alpha\sigma t} dt 
	= -\frac{1}{2\alpha\sigma} e^{-2\alpha\sigma t} \erf{t} + \frac{1}{2\alpha\sigma} e^{\alpha^2\sigma^2} \erf{t + \alpha\sigma} + C.
	\]
	
	Multiply by \(2\sigma\):
	\[
	\int \erf{\frac{r'}{2\sigma}  } e^{-\alpha r'} dr' 
	= -\frac{1}{\alpha} e^{-2\alpha\sigma t} \erf{t} + \frac{1}{\alpha} e^{\alpha^2\sigma^2} \erf{t + \alpha\sigma} + C.
	\]
	
	Substitute \(t = \frac{r'}{2\sigma}\):
	\[
	\int \erf{ \frac{r'}{2\sigma} } e^{-\alpha r'} dr' 
	= -\frac{1}{\alpha} e^{-\alpha r'} \erf{ \frac{r'}{2\sigma} } + \frac{1}{\alpha} e^{\alpha^2\sigma^2} \erf{ \frac{r'}{2\sigma} + \alpha\sigma} + C.
	\]
	
	\subsection{Evaluation of \(I_1(r)\)}
	
	\subsubsection*{First term}
	\begin{multline*}
		\int_0^r \erf{ \frac{r'}{2\sigma} } dr' 
		= \left[ r' \erf{ \frac{r'}{2\sigma} } + \frac{2\sigma}{\sqrt{\pi}} e^{-(r'/(2\sigma))^2} \right]_0^r \\
		= r \erf{ \frac{r}{2\sigma} } + \frac{2\sigma}{\sqrt{\pi}} \left( e^{-r^2/(4\sigma^2)} - 1 \right).
	\end{multline*}
	
	\subsubsection*{Second term} Using \(\alpha = 2k\):
	\begin{multline*}
		\int_0^r \erf{ \frac{r'}{2\sigma} } e^{-2k r'} dr' 
		= \left[ -\frac{1}{2k} e^{-2k r'} \erf{ \frac{r'}{2\sigma} } + \frac{1}{2k} e^{4k^2\sigma^2} \erf{ \frac{r'}{2\sigma} + 2k\sigma} \right]_0^r \\
		= -\frac{1}{2k} e^{-2k r} \erf{ \frac{r}{2\sigma} } + \frac{1}{2k} e^{4k^2\sigma^2} \left[ \erf{ \frac{r}{2\sigma} + 2k\sigma} - \erf{ 2k\sigma} \right].
	\end{multline*}
	
	Therefore, after simplification and grouping of terms one obtains
	\[
	I_1(r) = \left( \frac{r}{2}  	+ \frac{e^{-2k r} }{4k} \right) \erf{ \frac{r}{2\sigma} } + \frac{\sigma}{\sqrt{\pi}} \left( e^{-r^2/(4\sigma^2)} - 1 \right)
	- \frac{1}{4k} e^{4k^2\sigma^2} \left[ \erf{ \frac{r}{2\sigma} + 2k\sigma} - \erf{ 2k\sigma} \right].
	\]
	
	\subsection{Evaluation of \(I_2(r)\)}
	
	Using the indefinite integral with \(\alpha = 2k\):
	\[
	\int_r^\infty \erf{ \frac{r'}{2\sigma} } e^{-2k r'} dr' 
	= \left[ -\frac{1}{2k} e^{-2k r'} \erf{ \frac{r'}{2\sigma} } + \frac{1}{2k} e^{4k^2\sigma^2} \erf{ \frac{r'}{2\sigma} + 2k\sigma} \right]_r^\infty.
	\]
	
	As \(r' \to \infty\): \(\erf{ \frac{r'}{2\sigma} } \to 1\), \(\erf{ \frac{r'}{2\sigma} + 2k\sigma} \to 1\):
	\[
	\left[ -\frac{1}{2k} e^{-2k r'} \cdot 1 + \frac{1}{2k} e^{4k^2\sigma^2} \cdot 1 \right]_{r' \to \infty} = 0 + \frac{1}{2k} e^{4k^2\sigma^2}.
	\]
	
	At \(r' = r\):
	\[
	-\frac{1}{2k} e^{-2k r} \erf{ \frac{r}{2\sigma} } + \frac{1}{2k} e^{4k^2\sigma^2} \erf{ \frac{r}{2\sigma} + 2k\sigma}.
	\]
	
	Thus:
	\[
	I_2(r) = \frac{1}{2k} e^{4k^2\sigma^2} - \left[ -\frac{1}{2k} e^{-2k r} \erf{ \frac{r}{2\sigma} } + \frac{1}{2k} e^{4k^2\sigma^2} \erf{ \frac{r}{2\sigma} + 2k\sigma} \right].
	\]
	
	Simplify:
	\[
	I_2(r) = \frac{1}{2k} e^{4k^2\sigma^2} \left[ 1 - \erf{ \frac{r}{2\sigma} + 2k\sigma} \right] + \frac{1}{2k} e^{-2k r} \erf{ \frac{r}{2\sigma} }.
	\]
	
	\section{Proof of Lemma \ref{th:J3}}\label{sec:proofL1}
	
	The first inequality is obvious. Since $0\le y<1$ (at least on a set of positive
	measure), the advancing kernel satisfies
	\[
	J_2(0)=\int_0^\infty y(r')\,\frac{\erf{r'/2\sigma}}{r'}\,e^{-kr'}\,dr'
	<\int_0^\infty \frac{\erf{r'/2\sigma}}{r'}\,e^{-kr'}\,dr'=:J_3 .
	\]
	Set $a=1/2\sigma$. We show that $J_3$ has the closed form
	\begin{equation}\label{eq:J3closed}
		J_3(a,k)=\frac{2\zeta}{\sqrt{\pi}}\;{}_2F_2\!\left(\tfrac12,1;\tfrac32,\tfrac32;\zeta^2\right)
		-\frac12\,\operatorname{Ei}(\zeta^2),\qquad \zeta=\frac{k}{2a}=k\sigma .
	\end{equation}
	
	\subsection*{1. Mellin representation}
	
	Substituting $x=kr'$ and putting $\lambda:=a/k=1/(2\zeta)$,
	\[
	J_3=\int_0^\infty f(x)g(x)\,dx \qquad \text{with} \qquad
	f(x):=\frac{\erf{\lambda x}}{x},\quad g(x):=e^{-x}.
	\]
	With $\mathcal M\{h\}(s)=\int_0^\infty x^{s-1}h(x)\,dx$, Parseval's identity reads
	\[
	\int_0^\infty f g\,dx=\frac1{2\pi i}\int_{c-i\infty}^{c+i\infty}
	\mathcal M\{f\}(s)\,\mathcal M\{g\}(1-s)\,ds ,
	\]
	for $c$ in the common strip of analyticity.
	
	\begin{itemize}
		\item For $g$: $\mathcal M\{g\}(u)=\Gamma(u)$, $\Re u>0$, so $\mathcal M\{g\}(1-s)=\Gamma(1-s)$ for $\Re s<1$.
		\item For $f$: since $\erf{\lambda x}\sim 2\lambda x/\sqrt\pi$ at $0$ and $\to1$ at $\infty$,
		$\int_0^\infty x^{s-1}\erf{\lambda x}\,dx$ converges for $-1<\Re s<0$. Integrating by parts
		and using $\int_0^\infty x^{s}e^{-\lambda^2x^2}dx=\tfrac12\lambda^{-s-1}\Gamma\!\left(\tfrac{s+1}{2}\right)$,
		\[
		\int_0^\infty x^{s-1}\erf{\lambda x}\,dx
		=-\frac{2\lambda}{\sqrt\pi\,s}\int_0^\infty x^{s}e^{-\lambda^2x^2}dx
		=-\,\lambda^{-s}\,\frac{\Gamma\!\left(\frac{s+1}{2}\right)}{\sqrt{\pi}\,s},
		\qquad -1<\Re s<0 .
		\]
		Replacing $s\mapsto s-1$,
		\[
		\mathcal M\{f\}(s)=-\,\lambda^{1-s}\,\frac{\Gamma\!\left(\frac{s}{2}\right)}{\sqrt{\pi}\,(s-1)},
		\qquad 0<\Re s<1 .
		\]
	\end{itemize}
	The common strip is $0<\Re s<1$, and for any $c\in(0,1)$
	\begin{equation}\label{eq:mellinJ3}
		J_3=\frac1{2\pi i}\int_{c-i\infty}^{c+i\infty}
		\left[-\frac{\lambda^{1-s}\,\Gamma\!\left(\frac s2\right)\Gamma(1-s)}{\sqrt\pi\,(s-1)}\right]ds .
	\end{equation}
	Equivalently, the kernel can be written as
	$-\pi\lambda^{1-s}\Gamma\!\left(\frac{s-1}{2}\right)\big/\big[2^{s}\sin(\pi(1-s))\,\Gamma\!\left(\frac{s+1}{2}\right)^{2}\big]$,
	by the reflection and duplication formulas.
	
	\subsection*{2. Change of variable}
	
	Put $s=2z+1$, $ds=2\,dz$. Then $\lambda^{1-s}=\lambda^{-2z}$,
	$\Gamma(s/2)=\Gamma(z+\tfrac12)$, $\Gamma(1-s)=\Gamma(-2z)$, $s-1=2z$, and
	\begin{equation}\label{eq:Jz}
		J_3=-\frac1{\sqrt\pi}\,\frac1{2\pi i}\int_{\sigma_0-i\infty}^{\sigma_0+i\infty}
		F(z)\,dz,\qquad
		F(z):=\lambda^{-2z}\,\frac{\Gamma\!\left(z+\frac12\right)\Gamma(-2z)}{z}
	\end{equation}
	in the strip $ -\tfrac12<\sigma_0<0$.
	\subsection*{3. Residues}
	
	The singularities of $F$ to the right of the contour are:
	\begin{itemize}
		\item $z=0$: $\Gamma(z+\frac12)$ is regular, $\Gamma(-2z)$ has a simple pole and
		there is the factor $1/z$. Hence $z=0$ is a \emph{double} pole, which leads to a log-term in the development. 
		\item $z=n/2$, $n=1,2,\dots$: simple poles of $\Gamma(-2z)$. Since
		$\operatorname{Res}_{u=-n}\Gamma(u)=(-1)^n/n!$ and $u=-2z$,
		$\operatorname{Res}_{z=n/2}\Gamma(-2z)=-\dfrac{(-1)^n}{2\,n!}$.
	\end{itemize}
	Using $\Gamma(-2z)=\dfrac{-\pi}{2z\,\sin(2\pi z)\,\Gamma(2z)}$ and Stirling's formula,
	$|\Gamma(z+\tfrac12)\Gamma(-2z)|$ decays factorially along the arcs $|z|=R_N=N+\tfrac14$, $\Re z\ge\sigma_0$,
	so for every $\lambda>0$ the contour may be closed to the right.
	The closed contour is traversed clockwise, hence
	\begin{equation}\label{eq:sumres}
		J_3=\frac1{\sqrt\pi}\sum_{z_p\in\{0,\,1/2,\,1,\dots\}}\operatorname{Res}_{z=z_p}F(z).
	\end{equation}
	
	\paragraph{Simple poles.} For $n\ge1$,
	\[
	\operatorname{Res}_{z=n/2}F
	=-\frac{(-1)^n\,\lambda^{-n}\,\Gamma\!\left(\frac{n+1}{2}\right)}{n\;n!}.
	\]
	Recall $\lambda^{-1}=2\zeta$.
	
	\emph{Odd $n=2m+1$.} Here $\Gamma(m+1)=m!$ and
	$\dfrac{m!}{(2m+1)!}=\dfrac{1}{4^m(\frac32)_m}$, $\dfrac1{2m+1}=\dfrac{(\frac12)_m}{(\frac32)_m}$, so
	\[
	\frac1{\sqrt\pi}\sum_{m\ge0}\frac{(2\zeta)^{2m+1}\,m!}{(2m+1)\,(2m+1)!}
	=\frac{2\zeta}{\sqrt\pi}\sum_{m\ge0}\frac{(\frac12)_m(1)_m}{(\frac32)_m(\frac32)_m}\frac{\zeta^{2m}}{m!}
	=\frac{2\zeta}{\sqrt\pi}\,{}_2F_2\!\left(\tfrac12,1;\tfrac32,\tfrac32;\zeta^2\right).
	\]
	
	\emph{Even $n=2m$, $m\ge1$.} Using $\Gamma(m+\frac12)=\dfrac{(2m)!}{4^m m!}\sqrt\pi$ and $\lambda^{-2m}=4^m\zeta^{2m}$,
	\[
	\frac1{\sqrt\pi}\sum_{m\ge1}\left[-\frac{\lambda^{-2m}\,\Gamma(m+\frac12)}{2m\,(2m)!}\right]
	=-\frac12\sum_{m\ge1}\frac{\zeta^{2m}}{m\,m!}
	=-\frac12\Bigl[\operatorname{Ei}(\zeta^2)-\gamma-\ln\zeta^2\Bigr],
	\]
	because $\operatorname{Ei}(x)=\gamma+\ln x+\sum_{m\ge1}x^m/(m\,m!)$ for $x>0$.
	
	\paragraph{Double pole at $z=0$.} Write $F(z)=h(z)\,\Gamma(-2z)/z$ with
	$h(z)=\lambda^{-2z}\Gamma(z+\frac12)=\sqrt\pi\,\bigl[1+z\,(-2\ln\lambda+\psi(\tfrac12))+O(z^2)\bigr]$ and
	$\Gamma(-2z)=-\dfrac1{2z}-\gamma+O(z)$. Then
	\[
	\operatorname{Res}_{z=0}F=-\tfrac12h'(0)-\gamma h(0)
	=\sqrt\pi\Bigl[\ln\lambda+\tfrac{\gamma}{2}+\ln2-\gamma\Bigr]
	=\sqrt\pi\Bigl[-\ln\zeta-\tfrac{\gamma}{2}\Bigr],
	\]
	using $\psi(\tfrac12)=-\gamma-2\ln2$ and $\lambda=1/(2\zeta)$.
	
	\subsection*{4. Collecting terms}
	
	By \eqref{eq:sumres}, the constant and logarithmic terms cancel:
	\[
	\tfrac12\gamma+\tfrac12\ln\zeta^2-\ln\zeta-\tfrac{\gamma}{2}=0 ,
	\]
	and we obtain \eqref{eq:J3closed}:
	\[
	J_3(a,k)=\frac{2\zeta}{\sqrt{\pi}}\,{}_2F_2\!\left(\tfrac12,1;\tfrac32,\tfrac32;\zeta^2\right)
	-\tfrac12\operatorname{Ei}(\zeta^2),\qquad \zeta=\frac{k}{2a}. 
	\]
	
	\section{Implementation and detailed comparisons}\label{sec:numer}

	\subsection{Fredholm variational solvers}\label{sec:algorithms}
	
	\paragraph{Evaluation of the energy functional.}
	For a given decay parameter $k$ the first iterate $y_1(r;k)$ and its derivative are evaluated in
	closed form  on a grid $r\in[r_{\min},r_{\max}(k)]$ with
	$r_{\min}=10^{-6}$ and $r_{\max}=\max(20,10/k)$, using Simpson's rule with $n_r=2001$ points.
	The three integrals $N=\int y_1^2$, $T=\tfrac12\int (y_1')^2$ and $W=\int V y_1^2$ are
	vectorized, and $E_1(k)=(T+W)/N$. The terms containing
	$e^{x^2}\mathrm{erfc}(\cdot)$, $x=2k\sigma$, are evaluated through the scaled complementary error
	function,
	\[
	e^{x^2}\mathrm{erfc}(u+x)=\mathrm{erfcx}(u+x)\,e^{-u^2-2ux},\qquad u=\frac{r}{2\sigma},
	\]
	which is exact and overflow-free for all $k,\sigma$; no restriction such as $4k^2\sigma^2<25$ is
	needed. A node count of $y_1$ is retained as a sanity check ($y_1>0$ mathematically, since
	the kernel of $K_k$ is positive). Relative to a converged reference
	($n_r=2\times10^4$, Richardson-extrapolated), the energy differs by $\le1.5\times10^{-6}$ at
	$\sigma=0.1$ and by $\le10^{-7}$ for $\sigma\ge1$.
	
	\paragraph{Zeroth-order step.}
	The energy $E_0(k)$ of the starting function $y_0=re^{-kr}$ is available in closed form,
	$E_0(k)=k(2x^2-1)\,\mathrm{erfcx}(x)-2kx/\sqrt\pi+k^2/2$, $x=2k\sigma$. By
	Lemma~\ref{lem:lyapunov} it is strictly convex ($E_0''>1$) with $E_0'(0^+)=-1$ and
	$E_0'(k)>k-1$, so its minimizer $k_0^\ast\in(0,1)$ is the unique zero of $E_0'$ and is
	computed by Brent's root finder (6--17 evaluations, $|\Delta k|<10^{-9}$ against a bounded
	minimizer, for $\sigma\in[0.05,10^3]$).

	The step is halved until the energy decreases and grown by a factor $1.5$ after every success.
	Convergence is declared on the gradient, $|dE/dk|<\varepsilon_g$, not on $|E_{n+1}-E_n|$: with a
	fixed step the energy change per iteration is $\approx\eta g^2$, so a threshold on $|\Delta E|$
	stops far from the minimizer (for $\eta=0.05$, $|\Delta E|<10^{-5}$ leaves $k$ off by 1.5--4\%
	and $E_1$ too high by $7\times10^{-5}$--$10^{-4}$). The gradient is a central difference with
	$h_k=10^{-5}$, which balances the $O(h_k^2)$ truncation error against the quadrature noise
	divided by $h_k$. For $\sigma=0.1,\dots,20$ the algorithm converges in 12--53 iterations to
	$k_1=0.95457,\,0.71647,\,0.55451,\,0.23724,\,0.15347$ at $\sigma=0.1,0.5,1,5,10$, and these
	coincide with a Brent minimization of $E_1(k)$ over $[0.5k_0^\ast,\min(1,2k_0^\ast)]$ (10--12
	evaluations). In all tested cases $1.0<k_1/k_0^\ast<1.5$, so this bracket is sufficient.

	\paragraph{Higher iterates.}
	The second iterate is obtained from $y_1$ on the same grid. Writing
	$f=m\,y$ with $m(r)=\mathrm{erf}(r/2\sigma)/r$ and $m(0)=1/(\sigma\sqrt\pi)$,
	\[
	(K_ky)(r)=\frac1k\Bigl[P(r)+Q(r)-e^{-kr}B\Bigr]\]
	where
	\[
	P=\!\int_0^r\!e^{-k(r-t)}f\,dt,\;\; Q=\!\int_r^\infty\!e^{-k(t-r)}f\,dt,\;\; B=\!\int_0^\infty\! e^{-kt}f\,dt,
	\]
	and $P$, $Q$ obey one-step recursions in $r$ (forward for $P$, backward for $Q$), so $K_ky$
	costs $O(n_r)$ with no nested quadrature and no singular kernel. We checked
	that the closed-form $y_1$ and $K_ky_0$ agree to $\le7\times10^{-7}$ (relative, maximum norm).
	$E_2(k)$ is minimized over $k$ by Brent's method at each order; as shown in
	Remark~\ref{rem:kopt}, the fixed-$k$ second iterate is not an improvement for $\sigma\ge0.5$.

	\begin{table}[h]
		\centering
		\caption{Comparison of $k^*$ for $y_1$ by energy minimization 
			and by Virial Condition satisfaction ($\Delta_1=0$).}
		\begin{tabular}{lccc}
			\toprule
			$\sigma$ & $k_{\text{energy}}$ & $k_{\text{virial}}$ & Relative difference \\
			\midrule
			0.1 & 0.954568 & 0.953883 & $7.2\times10^{-4}$ \\
			0.5 & 0.716466 & 0.712845 & $5.1\times10^{-3}$ \\
			1.0 & 0.554512 & 0.551544 & $5.4\times10^{-3}$ \\
			5.0 & 0.237240 & 0.236999 & $1.0\times10^{-3}$ \\
			\bottomrule
		\end{tabular}
		\label{tab:virial-y1}
	\end{table}
	The small differences reported in Table~\ref{tab:virial-y1} reflect the fact that $y_1$, while more accurate than $y_0$, is still not the exact eigenfunction. 
	As the Fredholm iteration proceeds, one expects $\Delta_m[k_m^*] \to 0$, providing a useful 
	convergence criterion complementary to the energy itself.
	
	\subsection{Direct integration: shooting and Numerov}\label{sec:shooting-impl}
	
	For the reference solutions we integrate $y''=2\,(V_\sigma(r)-E)\,y$, $y(0)=0$, with
	$E=-k^2/2$. A priori brackets for the ground state are
	\begin{equation}\label{eq:bracket}
		\max\!\bigl(-\tfrac12,\,-\tfrac1{\sigma\sqrt\pi}\bigr)\;\le\;E_0\;\le\;\min\bigl(0,E_{\rm total}\bigr).
	\end{equation}
	The lower bound holds because $V_\sigma\ge-1/r$ (comparison with the hydrogen $s$-wave,
	$E=-\tfrac12$) and $V_\sigma\ge\min V_\sigma$; the upper bound is Lemma~\ref{lem:homomono}.

	Bisecting on the node count, not on the sign of $y(r_{\max})$, is essential: the sign of
	$y(r_{\max})$ changes at every eigenvalue and the bisection can converge to an excited level (we
	observed this at $\sigma=10$). The choice $r_{\max}=25/k$ makes the neglected growing mode
	irrelevant.

	The Numerov recurrence is
	$(1-\tfrac{h^2}{12}f_{i+1})y_{i+1}=2(1+\tfrac{5h^2}{12}f_i)y_i-(1-\tfrac{h^2}{12}f_{i-1})y_{i-1}$.
	The first sign change of $F$ scanning from $k=1$ downwards is the ground state, since $k_\ast<1$
	and lower roots are excited states. Rescaling the inward solution removes the poles that the
	log-derivative form has where $y(r_m)=0$. The decaying inward solution is the dominant one, so
	errors in its starting values are forgotten.
	
	\begin{table}[ht]
		\centering
		\caption{Ground-state energy (hartree): shooting, Numerov matching, and a second-order
			finite-difference diagonalization ($4\times10^4$ points).}\label{tab:direct}
		\begin{tabular}{lccc}
			\hline
			$\sigma$ & shooting (Alg.~\ref{alg:shoot}) & Numerov (Alg.~\ref{alg:numerov}) & finite difference \\
			\hline
			0.1 & $-0.47477539$ & $-0.47477539$ & $-0.47477566$ \\
			0.5 & $-0.33114072$ & $-0.33114072$ & $-0.33114077$ \\
			1   & $-0.23707528$ & $-0.23707528$ & $-0.23707530$ \\
			5   & $-0.07725572$ & $-0.07725572$ & $-0.07725573$ \\
			10  & $-0.04327087$ & $-0.04327087$ & $-0.04327087$ \\
			\hline
		\end{tabular}
	\end{table}
	
	Shooting and Numerov agree to $\lesssim10^{-12}$ at $\sigma=0.1,1,10$, and Numerov is unchanged
	to $10^{-13}$ on halving $h$. The residual deviation of the finite-difference values
	($2.7\times10^{-7}$ at $\sigma=0.1$) is the discretization error of the second-order
	diagonalization. The Numerov/shooting value is used as the reference energy throughout.
	
	\subsection{Numerical checks and cost}
	
	\paragraph{Grid convergence.}
	All quadratures are second order. For $E_1$ and $E_2$ at $k=k_1$, doubling the grid
	($n_r=1.5\times10^4,\,3\times10^4,\,6\times10^4$) changes the energy by $6.4\times10^{-7}$
	($\sigma=0.1$), $1.7\times10^{-7}$ ($0.5$), $7\times10^{-8}$ ($1$), $8\times10^{-9}$ ($5$) and
	$3\times10^{-9}$ ($10$) in the last refinement; Richardson extrapolation removes the leading
	error. Note that at $\sigma=0.1$ the grid error of $E_2^{\rm opt}$ at $n_r=3\times10^4$ is
	$\approx2.5\times10^{-6}$, comparable to its deviation from the reference energy ($4\times10^{-6}$),
	and extrapolated values should be quoted for $\sigma\le0.5$. The integration must start at
	$r=0$ (or include the interval $[0,r_{\min}]$): omitting $[0,0.01]$ removes
	$\tfrac12 y'(0)^2\cdot0.01$ from the kinetic energy and, with a 500-point grid, biases $E_1$ by
	$-1.3\times10^{-2}$ at $\sigma=0.1$ (below the exact ground-state energy, which is impossible for
	a Rayleigh quotient).
	
	\paragraph{Cost per energy evaluation.}
	One evaluation of $E_1(k)$ costs $O(n_r)$ vectorized operations ($\approx6$~ms at
	$n_r=2\times10^4$). A complete ground-state determination takes 12--53 gradient iterations
	($\approx3$--$4$ evaluations each) or 10--12 evaluations with Brent's method, i.e.\ 0.1--0.3~s.
	One evaluation of $F(k)$ in Alg.~\ref{alg:numerov} is one outward and one inward sweep,
	$O(R/h)$ operations; shooting needs 34--39 adaptive ODE solves. In our unoptimized pure-Python/SciPy
	implementation on one core a full determination took roughly 1--1.5~s for shooting and
	0.3--1~s for Numerov. For the single-particle radial ground state all three approaches are
	inexpensive; these timings are not performance claims about compiled implementations.
	
	
	\begin{algorithm}
		\caption{Fredholm gradient-descent ground state solver (adaptive step)}\label{alg:gd}
		\begin{algorithmic}[1]
			\Require $\sigma$; tolerance $\varepsilon_g=10^{-6}$; $h_k=10^{-5}$; $N_{\max}=200$;
			$\eta=0.05$, $\eta_{\max}=1.5$
			\State $k\gets k_0^\ast$ (zeroth-order minimizer); $E\gets E_1(k)$; $E_{\rm best}\gets\infty$
			\For{$n=1,\dots,N_{\max}$}
			\State \textbf{if} $E<E_{\rm best}$ \textbf{then} $(E_{\rm best},k_{\rm best})\gets(E,k)$
			\State $g\gets\bigl(E_1(k+h_k)-E_1(k-h_k)\bigr)/(2h_k)$
			\State \textbf{if} $|g|<\varepsilon_g$ \textbf{then break}
			\Repeat
			\State $k_{\rm new}\gets\min\bigl(k_{\max},\max(k_{\min},k-\eta g)\bigr)$; $E_{\rm new}\gets E_1(k_{\rm new})$
			\State \textbf{if} $E_{\rm new}\ge E$ \textbf{then} $\eta\gets\eta/2$
			\Until{$E_{\rm new}<E$}
			\State $k\gets k_{\rm new}$, $E\gets E_{\rm new}$, $\eta\gets\min(1.5\,\eta,\eta_{\max})$
			\EndFor
			\State \Return $k_1\gets k_{\rm best}$, $E_1\gets E_1(k_{\rm best})$
		\end{algorithmic}
	\end{algorithm}

	\begin{algorithm}
		\caption{Fredholm variational ground state: scan and refine (fallback)}\label{alg:scan}
		\begin{algorithmic}[1]
			\Require $\sigma$, $k_{\min}<k_{\max}\le1$, $n_{\rm scan}$
			\For{each $k$ in $\mathrm{linspace}(k_{\min},k_{\max},n_{\rm scan})$}
			\State $E[k]\gets E_1(k)$ and node count of $y_1$
			\State \textbf{if} $E[k]<0$ and nodes $=0$ \textbf{then} store as candidate
			\EndFor
			\State $k_s\gets\arg\min E[k]$ over candidates; refine on $[k_s-\Delta k,k_s+\Delta k]$ by Brent's method
			\State \Return $E_1=E(k^\ast)$, $k^\ast$ and the normalized $y_1$
		\end{algorithmic}
	\end{algorithm}
	
	\begin{algorithm}
		\caption{Outward shooting with node counting}\label{alg:shoot}
		\begin{algorithmic}[1]
			\Require $\sigma$, tolerance $\tau=10^{-12}$
			\State $E_{\rm lo}\gets\max(-\tfrac12,-1/(\sigma\sqrt\pi))$; $E_{\rm hi}\gets\min(0,E_{\rm total}(\sigma))$
			\While{$E_{\rm hi}-E_{\rm lo}>\tau$}
			\State $E\gets\tfrac12(E_{\rm lo}+E_{\rm hi})$, $k\gets\sqrt{-2E}$, $r_{\max}\gets25/k$
			\State Integrate from $r_0=10^{-8}$ with $y(r_0)=r_0$, $y'(r_0)=1$ (DOP853, rtol $=10^{-12}$)
			\State $n\gets$ number of sign changes of $y$ on $(0,r_{\max})$
			\If{$n=0$} $E_{\rm lo}\gets E$ \Else\ $E_{\rm hi}\gets E$ \EndIf
			\EndWhile
			\State \Return $E_0\approx\tfrac12(E_{\rm lo}+E_{\rm hi})$
		\end{algorithmic}
	\end{algorithm}
	
	\begin{algorithm}
		\caption{Two-sided Numerov matching}\label{alg:numerov}
		\begin{algorithmic}[1]
			\Require $\sigma$, step $h$ ($=2\times10^{-3}$), matching point $r_m=1+\sigma$
			\For{$k$ from $1$ downwards on a grid}
			\State $R\gets30/k$
			\State Outward: $y_{\rm out}(0)=0$, $y_{\rm out}(h)=h$; inward: $y_{\rm in}(R)=1$, $y_{\rm in}(R-h)=e^{kh}$,
			Numerov recurrence with $f=2(V_\sigma+k^2/2)$
			\State Rescale $y_{\rm in}\gets y_{\rm in}\,y_{\rm out}(r_m)/y_{\rm in}(r_m)$
			\State $F(k)\gets y_{\rm out}'(r_m)-y_{\rm in}'(r_m)$ (central differences)
			\If{$F$ changes sign between consecutive $k$} refine the root by Brent's method; \Return \EndIf
			\EndFor
		\end{algorithmic}
	\end{algorithm}

	
	\begin{table}[h!]
	
	Comparisons for $\sigma=0.1$
		
	\begin{tabular}{lccccc}\toprule
		& exact & $y_0$ & $y_1$ & $y_2(k_1)$ & $y_2(k_2)$ \\ \midrule
		$E$ & -0.474775 & -0.472933 & -0.474709 & -0.474723 & -0.474770 \\
		$k$ & 0.9744 & 0.9277 & 0.9546 & 0.9546 & 0.9647 \\
		$\langle r\rangle$ & 1.579 & 1.617 & 1.587 & 1.598 & 1.582 \\
		$\sqrt{\langle r^2\rangle}$ & 1.816 & 1.867 & 1.829 & 1.839 & 1.820 \\
		nodes & 0 & 0 & 0 & 0 & 0 \\
		$E/E_{\min}$ & 0.0842 & 0.0838 & 0.0841 & 0.0841 & 0.0842 \\
		\bottomrule\end{tabular}  
	
	Comparisons for $\sigma=0.5$
	
	\begin{tabular}{lccccc}\toprule
		& exact & $y_0$ & $y_1$ & $y_2(k_1)$ & $y_2(k_2)$ \\ \midrule
		$E$ & -0.331141 & -0.321730 & -0.330458 & -0.330258 & -0.331072 \\
		$k$ & 0.8138 & 0.6120 & 0.7165 & 0.7165 & 0.7626 \\
		$\langle r\rangle$ & 2.234 & 2.451 & 2.274 & 2.367 & 2.245 \\
		$\sqrt{\langle r^2\rangle}$ & 2.516 & 2.830 & 2.584 & 2.678 & 2.537 \\
		nodes & 0 & 0 & 0 & 0 & 0 \\
		$E/E_{\min}$ & 0.2935 & 0.2851 & 0.2929 & 0.2927 & 0.2934 \\
		\bottomrule\end{tabular}  
	
	Comparisons for $\sigma=1.0$
	
	\begin{tabular}{lccccc}\toprule
		& exact & $y_0$ & $y_1$ & $y_2(k_1)$ & $y_2(k_2)$ \\ \midrule
		$E$ & -0.237075 & -0.228396 & -0.236181 & -0.235789 & -0.236947 \\
		$k$ & 0.6886 & 0.4450 & 0.5545 & 0.5545 & 0.6113 \\
		$\langle r\rangle$ & 3.035 & 3.370 & 3.101 & 3.302 & 3.055 \\
		$\sqrt{\langle r^2\rangle}$ & 3.379 & 3.892 & 3.499 & 3.703 & 3.419 \\
		nodes & 0 & 0 & 0 & 0 & 0 \\
		$E/E_{\min}$ & 0.4202 & 0.4048 & 0.4186 & 0.4179 & 0.4200 \\
		\bottomrule\end{tabular}  
	
	Comparisons for $\sigma=5.0$
	
	\begin{tabular}{lccccc}\toprule
		& exact & $y_0$ & $y_1$ & $y_2(k_1)$ & $y_2(k_2)$ \\ \midrule
		$E$ & -0.077256 & -0.074670 & -0.076775 & -0.076429 & -0.077120 \\
		$k$ & 0.3931 & 0.1741 & 0.2372 & 0.2372 & 0.2791 \\
		$\langle r\rangle$ & 7.928 & 8.614 & 8.081 & 9.083 & 7.981 \\
		$\sqrt{\langle r^2\rangle}$ & 8.685 & 9.946 & 9.052 & 10.076 & 8.844 \\
		nodes & 0 & 0 & 0 & 0 & 0 \\
		$E/E_{\min}$ & 0.6847 & 0.6617 & 0.6804 & 0.6773 & 0.6835 \\
		\bottomrule\end{tabular}  
	
	Comparisons for $\sigma=10.0$
	
	\begin{tabular}{lccccc}\toprule
		& exact & $y_0$ & $y_1$ & $y_2(k_1)$ & $y_2(k_2)$ \\ \midrule
		$E$ & -0.043271 & -0.042074 & -0.043011 & -0.042796 & -0.043184 \\
		$k$ & 0.2942 & 0.1104 & 0.1535 & 0.1535 & 0.1835 \\
		$\langle r\rangle$ & 12.695 & 13.585 & 12.888 & 14.782 & 12.758 \\
		$\sqrt{\langle r^2\rangle}$ & 13.862 & 15.686 & 14.430 & 16.375 & 14.127 \\
		nodes & 0 & 0 & 0 & 0 & 0 \\
		$E/E_{\min}$ & 0.7670 & 0.7457 & 0.7624 & 0.7585 & 0.7654 \\
		\bottomrule\end{tabular} 
	\caption{Detailed comparisons}
	\end{table}
	
\end{document}